\documentclass[aps,prd,preprint,amsmath,amssymb,nofootinbib,eqsecnum,superscriptaddress]{revtex4-2}
\usepackage[colorlinks=true,linkcolor=blue,citecolor=blue,urlcolor=blue]{hyperref}

\allowdisplaybreaks 

\begin{document}

\title{Regularization of Gauss--Bonnet Gravity in Riemann--Cartan Geometry}

\author{Jianhui Qiu}
\email{jhqiu@ucas.ac.cn}
\affiliation{School of Fundamental Physics and Mathematical Sciences,
Hangzhou Institute for Advanced Study, UCAS, Hangzhou 310024, China}
\author{Ling-Wei Luo}
\email{lwluo.phys@gmail.com}
\affiliation{School of Fundamental Physics and Mathematical Sciences,
Hangzhou Institute for Advanced Study, UCAS, Hangzhou 310024, China}
\author{Chunhui Liu}
\email{liuchunhui22@mails.ucas.ac.cn}
\affiliation{School of Fundamental Physics and Mathematical Sciences,
Hangzhou Institute for Advanced Study, UCAS, Hangzhou 310024, China}
\affiliation{University of Chinese Academy of Sciences, Beijing 100190, China}
\affiliation{Institute of Theoretical Physics, Chinese Academy of Sciences, Beijing 100190, China}
\author{Chao-Qiang Geng}
\email{cqgeng@ucas.ac.cn}
\affiliation{School of Fundamental Physics and Mathematical Sciences,
Hangzhou Institute for Advanced Study, UCAS, Hangzhou 310024, China}
\affiliation{Department of Physics and Synergetic Innovation Center for Quantum Effects and Applications,
Hunan Normal University, Changsha, Hunan 410081, China}


\begin{abstract}
We extend the conformal dimensional-derivative regularization of four-dimensional Gauss--Bonnet gravity to Riemann--Cartan geometry, obtaining a regularized action whose torsionless limit equals the well-known regularized four-dimensional Einstein--Gauss--Bonnet model. Varying independently with respect to the scalar, tetrad, and spin connection yields field equations that remain strictly second order in covariant derivatives, thereby avoiding Ostrogradsky-type instabilities. Within this framework we obtain static, spherically symmetric black holes carrying torsion hair, showing that the regularized Gauss--Bonnet interaction can support long-range torsion hair without invoking extra dimensions.
\end{abstract}

\maketitle
\newpage

\section{Introduction}

As one of the fundamental interactions in nature, gravity is described in its contemporary form by  General Relativity (GR) of Einstein, which is based on Riemannian geometry with a torsion-free and metric-compatible Levi--Civita connection. However, unresolved issues, such as the singularity problem, the origin of dark energy and dark matter, and the early-universe inflation suggest that a systematic extension of GR is necessary.

Higher-order curvature terms naturally arise in modifications of gravitational theory. From the perspective of effective field theory, higher-order curvature corrections must appear in the low-energy effective action of GR \cite{burgess2004quantum}. Meanwhile, gravity with quadratic curvature terms is perturbatively renormalizable, although it is plagued by the ghost problem \cite{stelle1977renormalization}. Furthermore, the $R^2$-driven Starobinsky inflation model agrees well with observations \cite{starobinsky1980new, planck2018results}. In higher-dimensional spacetimes, Lanczos--Lovelock theory provides the most general geometric extension that preserves second-order field equations, leading to testable new effects in black hole and holographic thermodynamics, such as the generalization of the Iyer--Wald entropy formula, corrections to the viscosity-to-entropy ratio ($\eta/s$) bound, and the calculation of holographic entanglement entropy in higher-derivative gravity \cite{padmanabhan2013lanczos, iyer1994some, brigante2008viscosity, dong2014holographic}.

Meanwhile, most quantum fields carry intrinsic spin. Just as the energy-momentum tensor of matter couples to the spacetime metric, the spin density of matter should naturally couple to a geometric quantity of spacetime—torsion. This leads to Riemann–Cartan (RC) geometry and the Poincaré gauge theory (PGT) of gravity built upon it. In its minimal model, the Einstein--Cartan--Sciama--Kibble (ECSK) theory, or simply the Einstein--Cartan (EC) theory, both the energy--momentum tensor and the spin density act as gravitational sources. This yields an effective repulsive force from the spin-torsion coupling in high-density regions, which helps to alleviate the singularity problem. However, the field equation for torsion in this theory is an algebraic constraint that directly ties torsion to the spin density of matter \cite{hehl1976general, hehl1995metric, blagojevic2012gauge}. A direct consequence is that, in vacuum regions {without spin}, torsion must vanish. This severely limits the possibility of exploring torsional effects generated by purely geometric structures.

To overcome this limitation of EC theory, a natural extension is to introduce higher-order curvature terms into {the model}, thus forming Lovelock--Cartan theory. Within this framework, the algebraic constraint on torsion is {replaced} by {differential equations in the} presence of higher-order curvature terms, {for example, the so-called Gauss--Bonnet (GB) term.} Torsion is no longer solely excited by matter spin but can be induced by the higher-order curvature terms themselves \cite{castellani2017palatini, baekler2011gravity}. This change fundamentally opens the door to find nontrivial vacuum solutions—even in the absence of matter, a dynamic curvature could self-consistently generate nonzero torsion. Previous studies have successfully constructed such curvature--induced black hole solutions with torsion in higher dimensions, particularly in five-dimensional {Einstein--Cartan--Gauss--Bonnet (5DECGB)} model in Lovelock gravity \cite{cvetkovic2018black}, as well as other exact solutions under special parameter choices \cite{canfora2014exact, oliva2011static}. These works reveal that torsion can act as an effective charge in higher-dimensional spacetimes, significantly impacting properties such as black hole thermodynamics.

Our focus is {on} the  GB term, the leading nontrivial higher-curvature correction to Einstein gravity. In a four-dimensional spacetime with pure Riemannian geometry, the GB term is a topological invariant, which does not contribute to the bulk dynamical equations \cite{padmanabhan2013lanczos}. This conclusion remains valid in 4D RC geometry \cite{nieh1980gauss,iosifidis2021riemann}, thus requiring a specific regularization scheme to induce effective four-dimensional dynamics. In the torsion-free case, the main approaches to make it dynamical include: (i) taking the dimensional limit $D\to4$ with a rescaling of the coupling constant \cite{glavan2020einstein}; (ii) demonstrating the equivalence of 4D Einstein--Gauss--Bonnet (EGB) {theory} to specific scalar-tensor theories (of the Horndeski/DHOST class) through Kaluza–Klein reduction or field redefinitions \cite{lu2020horndeski, kobayashi2020effective};   and (iii)  introducing a conformal regularization method with an auxiliary scalar field \cite{fernandes20224d,fernandes2020derivation}.
For these torsion-free 4D EGB realizations, the resulting phenomenology has been widely explored. For instance, the black hole thermodynamics exhibits a phase structure analogous to a Van der Waals fluid in extended phase space \cite{wei2020extended}; studies on the stability and quasinormal modes of the solutions indicate potential instabilities in certain parameter ranges \cite{fernandes20224d}; and in cosmology, the Friedmann–Lemaître–Robertson–Walker  background solutions and their linear perturbations have been analyzed, with observational constraints on the theory's parameters derived from cosmological data \cite{haghani2020growth, gohain2024emergent, khodabakhshi2024observational, clifton2020observational}. 

Although various classes of exact solutions with torsion and their thermodynamic analyses exist in higher-dimensional Lovelock--Cartan theories \cite{cvetkovic2019entropy, blagojevic2022entropy, blagojevic2006black}, a regularization scheme and phenomenological framework for {ECGB} theory truly aimed at four-dimensional spacetime, which can be compared in parallel with the well-established 4D EGB theory, remains to be established. 
In this work we extend the conformal (dimensional-derivative) regularization of four-dimensional Einstein--Gauss--Bonnet gravity to Riemann--Cartan geometry and construct a regularized Einstein--Cartan--Gauss--Bonnet (rECGB) action whose torsionless sector reproduces a scalar--tensor realization of 4D Einstein--Gauss--Bonnet gravity, and can therefore be interpreted as its Riemann--Cartan \cite{fernandes20224d,fernandes2020derivation} generalization. A key structural result is that, although the regularized Gauss--Bonnet density contains second derivatives of the scalar and explicit torsion, all Euler--Lagrange equations obtained by varying independently with respect to the scalar, tetrad and spin connection remain strictly second order in covariant derivatives of the dynamical fields, so that no Ostrogradsky-type higher-derivative ghost is introduced. On this basis we construct static, spherically symmetric black hole solutions endowed with long-range torsion hair, and show that the regularized Gauss--Bonnet interaction can act as an intrinsic geometric source for torsion already in four dimensions, providing a purely four-dimensional mechanism to sustain torsion hair without invoking extra dimensions.

The paper is organized as follows. In Section~II we construct the conformally regularized  GB action in  RC geometry. Section~III derives the full field equations by variation with respect to the scalar, tetrad, and spin connection. In Section~IV we present static, spherically symmetric black hole solutions with torsion hair under different ansätze and analyze their geometric properties. Section~V concludes. Technical details are collected in Appendices.

Throughout this paper we adopt the following  conventions. Greek indices
$\mu,\nu,\dots$ label spacetime components in a coordinate basis, whereas Latin
indices $a,b,\dots$ label components in a local Lorentz (orthonormal tetrad)
frame. A ring accent indicates torsionless (Levi–Civita) objects built from
$g_{\mu\nu}$, e.g. $\mathring{\nabla}_\mu$, $\mathring{\Gamma}{}^{\lambda}{}_{\mu\nu}$, and
$\mathring{R}{}^{\alpha}{}_{\beta\mu\nu}$. Symbols without the ring, such as
$\nabla_\mu$ or $R^{\alpha}{}_{\beta\mu\nu}$, are constructed from the full
RC connection with torsion. A tilde denotes quantities in the
conformally rescaled frame, e.g. $\tilde g_{\mu\nu}=e^{2\phi} g_{\mu\nu}$ with the
associated $\tilde{\nabla}_\mu$ and $\tilde R^{\alpha}{}_{\beta\mu\nu}$.

\section{Regularization of the action}

The Lagrangian density for the GB term depends on the tetrad one-form $\boldsymbol{e}$ and the spin connection one-form $\boldsymbol{\omega}$. In exterior calculus, using the curvature two-form,
\begin{equation}
  R^{ij} = d\omega^{ij} + \omega^{ik}\wedge\omega_k{}^{\,j}
         = \frac{1}{2}\,R^{ij}{}_{\mu\nu}\,dx^\mu\wedge dx^\nu \,, 
\end{equation}
the Lagrangian $n$-form can be written as \cite{mardones1991lovelock}
\begin{equation}
  \mathcal{L}_{GB}
  = \varepsilon_{i_1 i_2 \cdots i_n}\,
    R^{i_1 i_2}\wedge R^{i_{3} i_{4}}\wedge
    e^{i_{5}}\wedge \cdots \wedge e^{i_n} \,,
\end{equation}
where, $\varepsilon_{i_1 i_2 \ldots i_n}$ denotes the Levi--Civita symbol with the convention $\varepsilon_{0\,1\,\cdots\,n-1}=1$.

The contribution of $\mathcal{L}_{GB}$ to the action takes the form
\begin{equation}
\frac{(n-4)!}{4}\,\sqrt{-g}\,\delta_{\nu_1 \ldots \nu_4}^{\mu_1 \ldots \mu_4}\,
R^{\nu_1\nu_2}{}_{\mu_1\mu_2}\,R^{\nu_3\nu_4}{}_{\mu_3\mu_4} \,.
\end{equation}
It is therefore conventional to normalize through dividing $\mathcal{L}_{GB}$ by the factorial $(n-4)!$. This yields the standard form 
\begin{equation}\label{Eq:GB with torsion}
\frac{1}{4}\,\sqrt{-g}\,\delta_{\nu_1 \ldots \nu_4}^{\mu_1 \ldots \mu_4}\,
R^{\nu_1\nu_2}{}_{\mu_1\mu_2}\,R^{\nu_3\nu_4}{}_{\mu_3\mu_4}
= \sqrt{-g}\!\left(
R^{\alpha\beta\mu\nu}R_{\mu\nu\alpha\beta}
-4R^{\nu}{}_{\alpha}{}^{\mu\alpha}R_{\mu\beta\nu}{}^{\beta}
+R^{\mu\beta}{}_{\mu\beta}R^{\nu\alpha}{}_{\nu\alpha}
\right) \,.
\end{equation}

For any tetrad $\boldsymbol{e}$, one introduces the  Levi--Civita {spin} connection:
\begin{equation}
    \mathring\omega_{ab\mu}[\boldsymbol{e}]
= e_{[a}{}^{\rho} e_{b]}{}^{\sigma}\!\left(
  e_{c\sigma}\,\partial_{\mu} e^{c}{}_{\rho}
  + \partial_{\sigma}\!\big(e^{c}{}_{\rho}\,e_{c\mu}\big)
\right)
\end{equation}
Under the conformal rescaling of the tetrad, $\boldsymbol{e}\to e^{\phi}\boldsymbol{e}$, the Levi–Civita spin connection transforms as:
\begin{equation}
\mathring\omega{}^{a}{}_{b\mu}\to \mathring\omega{}^{a}{}_{b\mu}
- e_{b\mu}e^{a\nu}\partial_{\nu}\phi + e_{a\mu}e^{b\nu}\partial_{\nu}\phi \,,
\end{equation}
where the internal metric ${\eta_{ab}}=\mathrm{diag}(-,+,+,+)$ is used to raise and lower Lorentz indices. A general spin connection can always be decomposed into the Levi--Civita {spin} connection and the contortion tensor {with the expression of}
\begin{equation}
 \omega{}^{a}{}_{b\mu}= \mathring\omega{}^{a}{}_{b\mu}+K^{a}{}_{b\mu} \,,
\end{equation}
and its curvature decomposes as:
\begin{equation}
   R^{a}{}_{b}= \mathring{R}{}^{a}{}_{b}
   + d_{\mathring\omega{}}(K^{a}{}_{b}) + K^{a}{}_{c}\wedge K^{c}{}_{b} \,,
\end{equation}
where $d_{\mathring\omega{}}$ is the exterior covariant differentiation with respect to the Levi--Civita connection.

To determine how the curvature transforms, we first fix the conformal transformation of the connection. Within the framework of theories with torsion, two principal schemes for the conformal transformation of the spin connection have been identified; see also Ref.~\cite{iosifidis2019scale} for a systematic classification of scale (frame and connection) transformations in metric--affine geometry. In the first scheme, due to Nieh and Yan \cite{nieh1982quantized}, the spin connection itself is conformally invariant. In the second scheme—which we adopt—the contortion tensor is conformally invariant \cite{chakrabarty2018different}. The latter approach ensures that in the torsion-free limit, the transformation rules for all geometric entities are reduced to their standard forms in conventional {GR}, thereby providing a natural generalization of the Riemannian case.

Our choice of the conformally invariant contortion,
\begin{equation}
    K^{a}{}_{b} \to  K^{a}{}_{b} \,,  
\end{equation}
is primarily motivated by the requirement that the conformal behavior of the GB term must be correctly reduced to its well-established form in pure Riemannian geometry when torsion vanishes. It then follows that the general spin connection transforms as:
\begin{equation}
\omega{}^{a}{}_{b\mu}\to \omega{}^{a}{}_{b\mu}
- e_{b\mu}e^{a\nu}\partial_{\nu}\phi + e^{a}{}_{\mu}e_b{}^{\nu}\partial_{\nu}\phi \,.
\end{equation}
Transitioning to the coordinate basis, the affine connection is related to its spin counterpart via
\begin{equation}
  \Gamma^\rho{}_{\nu \mu}
  = e_a{}^\rho \partial_\mu e^a{}_\nu + e_a{}^\rho \omega ^a{}_{b \mu} e^b{}_\nu \,,
\end{equation}
which admits the decomposition
\begin{equation}
  \Gamma^\rho{}_{\nu \mu}= \mathring\Gamma{}^\rho{}_{\nu \mu}+K^\rho{}_{\nu \mu} \,,
\end{equation}
{and then torsion tensor is given by
\begin{equation}
T^\rho{}_{\mu \nu} = \Gamma^\rho{}_{\nu \mu} - \Gamma^\rho{}_{\mu \nu} =  K^\rho{}_{\nu \mu} - K^\rho{}_{\mu \nu} \,.
\end{equation}}%
Under a conformal transformation, the Levi--Civita connection transforms as
\begin{equation}
\mathring\Gamma{}^\lambda{}_{\alpha \beta}\to
\mathring\Gamma{}^\lambda{}_{\alpha \beta}
+\delta_\alpha^\lambda \mathring{\nabla}_\beta \phi
+\delta_\beta^\lambda \mathring{\nabla}_\alpha \phi
- g_{\alpha \beta} \mathring{\nabla}{}^\lambda \phi \,, 
\end{equation}
where $\mathring{\nabla}_\beta \phi=\partial_{\beta}\phi$. Since the contortion is assumed {to be} invariant, the transformation of the full connection is given by
\begin{equation}
\Gamma{}^\lambda{}_{\alpha \beta}\to
\Gamma{}^\lambda{}_{\alpha \beta}
+\delta_\alpha^\lambda {\nabla}_\beta \phi
+\delta_\beta^\lambda {\nabla}_\alpha \phi
- g_{\alpha \beta} {\nabla}{}^\lambda \phi \,.
\end{equation}
The Riemann curvature tensor in a coordinate basis,
\begin{equation}
    R^{\alpha}{}_{\beta\mu\nu}
    = \partial_{\mu}\Gamma^{\alpha}{}_{\beta\nu}
     - \partial_{\nu}\Gamma^{\alpha}{}_{\beta\mu}
     + \Gamma^{\alpha}{}_{\lambda\mu}\Gamma^{\lambda}{}_{\beta\nu}
     - \Gamma^{\alpha}{}_{\lambda\nu}\Gamma^{\lambda}{}_{\beta\mu} \,,
\end{equation}
consequently transforms as
\begin{equation}
R^{\alpha}{}_{\beta\mu\nu} \to  R^{\alpha}{}_{\beta\mu\nu}
+ \nabla_{\mu}\delta\Gamma^{\alpha}{}_{\beta\nu}
- \nabla_{\nu}\delta\Gamma^{\alpha}{}_{\beta\mu}
+ \delta\Gamma^{\alpha}{}_{\lambda\mu}\delta\Gamma^{\lambda}{}_{\beta\nu}
- \delta\Gamma^{\alpha}{}_{\lambda\nu}\delta\Gamma^{\lambda}{}_{\beta\mu}
+ T^{\lambda}{}_{\mu\nu}\delta\Gamma^{\alpha}{}_{\beta\lambda} \,.
\end{equation}
Substituting the expression for $\delta\Gamma{}^\lambda{}_{\alpha \beta}
{:=} \delta_\alpha^\lambda {\nabla}_\beta \phi+\delta_\beta^\lambda {\nabla}_\alpha \phi
- g_{\alpha \beta} {\nabla}{}^\lambda \phi$, we have
\begin{align*}
R^{\alpha}{}_{\beta \mu \nu} \to\; & R^{\alpha}{}_{\beta \mu \nu}
- T_{\beta \mu \nu}\,(\nabla^\alpha \phi)
+ T^{\alpha}{}_{\mu \nu}\,(\nabla_\beta \phi)
+ \delta^{\alpha}_{\nu} g_{\beta \mu} (\nabla_\gamma \phi) (\nabla^\gamma \phi)
- \delta^{\alpha}_{\mu} g_{\beta \nu} (\nabla_\gamma \phi) (\nabla^\gamma \phi) \\
& + g_{\beta \nu} (\nabla^\alpha \phi) (\nabla_\mu \phi)
- \delta^{\alpha}_{\nu} (\nabla_\beta \phi) (\nabla_\mu \phi)
- g_{\beta \nu} \nabla_\mu (\nabla^\alpha \phi)
+ \delta^{\alpha}_{\nu} \nabla_\mu (\nabla_\beta \phi) \\
& - g_{\beta \mu} (\nabla^\alpha \phi) (\nabla_\nu \phi)
+ \delta^{\alpha}_{\mu} (\nabla_\beta \phi) (\nabla_\nu \phi)
+ g_{\beta \mu} \nabla_\nu (\nabla^\alpha \phi)
- \delta^{\alpha}_{\mu} \nabla_\nu (\nabla_\beta \phi) \,.
\end{align*}
Accordingly, the GB term
$\mathcal{G}=R^{\alpha\beta\mu\nu}R_{\mu\nu\alpha\beta}
-4R^{\nu}{}_{\alpha}{}^{\mu\alpha}R_{\mu\beta\nu}{}^{\beta}
+R^{\mu\beta}{}_{\mu\beta}R^{\nu\alpha}{}_{\nu\alpha}$
transforms as 
\begin{equation}
\begin{aligned}
\sqrt{-g}\,\mathcal{G} \to \sqrt{-\tilde{g}}\,\tilde{\mathcal{G}}
= {} & \sqrt{-g}\, e^{(n-4)\phi} \Big\{ \mathcal{G}
+ 8 (n - 3) R^{\alpha\beta} \big(\nabla_{\alpha} \nabla_{\beta} \phi
- \nabla_{\alpha} \phi \nabla_{\beta} \phi\big) \\
& - 2 (n - 3) (n - 4) R (\nabla \phi)^2
+ 4 (n - 2) (n - 3)^2 (\nabla \phi)^2\,(\nabla^{2} \phi) \\
& - 4 (n - 2) (n - 3)\,(\nabla_{\alpha} \nabla_{\beta} \phi)
(\nabla^{\alpha} \nabla^{\beta} \phi)
+ 4 (n - 2) (n - 3) (\nabla^{2} \phi)^2 \\
& + 8 (n - 2) (n - 3)\,(\nabla^{\beta} \nabla^{\alpha} \phi)
(\nabla_{\alpha} \phi) (\nabla_{\beta} \phi)
- 4 (n - 3) R (\nabla^{2} \phi) \\
& + (n - 1) (n - 2) (n - 3) (n - 4) (\nabla \phi)^4 \\
& + 8(n-3)\!\left[T_{\alpha}(\nabla^{\alpha}\phi)(\nabla^2\phi)
- T_{\beta}(\nabla_{\alpha}\nabla^{\beta}\phi)\nabla^{\alpha}\phi
- T_{\mu\alpha\beta}\nabla^{\alpha}\phi (\nabla^{\mu}\nabla^{\beta}\phi)\right] \\
& + 4(n-2)(n-3)\!\left[T_{\alpha}\nabla^{\alpha}\phi (\nabla\phi)^2
+ T_{\alpha\beta\mu}\nabla^{\alpha}\phi(\nabla^{\mu}\nabla^{\beta}\phi)\right] \\
& + 8T^{\beta}{}_{\alpha}{}^{\mu}R_{\mu\beta}\nabla^{\alpha}\phi
+ 8T^{\mu}R_{\mu\alpha}\nabla^{\alpha}\phi
- 4T^{\beta\mu\nu}R_{\mu\nu\alpha\beta}\nabla^{\alpha}\phi
- 4T_{\alpha}R\nabla^{\alpha}\phi \Big\} \,.
\end{aligned}
\end{equation}

Within this conformal regularization framework, two distinct approaches are available. The first approach takes the action
\begin{equation}
S_1 = \lim_{n \to 4} \left[
\int d^n x \,\sqrt{-g}\,\left( \frac{\alpha}{n-4}\,\mathcal{G} \right)
- \frac{\alpha}{n-4} \int d^n x \,\sqrt{-\tilde{g}}\, \tilde{\mathcal{G}}
\right] \,.
\end{equation}
In this approach, a conformally transformed counterterm is subtracted before taking the limit $n\to4$. The procedure relies on identifying and discarding terms that diverge in this limit but qualify as boundary terms, thereby yielding a finite result.

The second method, which we adopt, employs a dimensional derivative, which is defined by
\begin{equation}
S_2 = \alpha \lim _{n \rightarrow 4}
\frac{ \int d^n x \sqrt{-\tilde{g}}\, \tilde{\mathcal{G}}
- \int d^4 x \sqrt{-\tilde{g}}\, \tilde{\mathcal{G}} }{n - 4} \,.
\end{equation}
This is equivalent to evaluating
\begin{equation}
S_2 = \alpha \lim _{n \rightarrow 4}
\int d^n x \,\frac{d}{d n} \!\left( \sqrt{-\tilde{g}}\, \tilde{\mathcal{G}} \right) \,.
\end{equation}
This technique differs fundamentally from the first, as it incorporates a counter-term whose numerator is evaluated directly in the target dimension of four dimensions (within the conformally transformed frame). A significant advantage is that it inherently avoids introducing a divergent prefactor like $\frac{1}{n-4}$ in the final limit.

In GR, the two schemes produce identical effective actions. However, it should be noted that this equivalence breaks down in RC geometry, where torsion is present. The first method relies on discarding the divergent pieces as boundary contributions. This procedure is justified only when working with the Levi--Civita connection, where the identity
\begin{equation}
\int d^n x \sqrt{-g}\, \mathring{\nabla}_\mu V^\mu = \text{boundary term} \,,
\end{equation}
holds. In contrast, for a general connection with torsion, the covariant derivative $\nabla_\mu V^\mu$ does not generally yield a pure boundary term:
\begin{equation}
\int d^n x \sqrt{-g}\, \nabla_\mu V^\mu \neq \text{boundary term} \,.
\end{equation}
Consequently, in RC geometry those divergent terms cannot, in general, be discarded as boundary contributions, rendering the scheme ambiguous. By contrast, the second method avoids discarding any terms and thus yields a consistent, well-defined regularization.

Applying this conformal regularization on the GB term gives
\begin{equation}
\label{fullaction1}
\begin{aligned}
\left.\frac{d}{dn}\!\big( \sqrt{-\tilde{g}}\, \tilde{\mathcal{G}}\big)\right|_{n=4}
= {} & \sqrt{-g}\,\Big\{ \phi\,\mathcal{G}
+ 8 (1+\phi) R^{\alpha\beta} \big(\nabla_{\alpha} \nabla_{\beta} \phi
- \nabla_{\alpha} \phi \nabla_{\beta} \phi\big) \\
& - 2 R (\nabla \phi)^2
+ (20+8\phi)(\nabla \phi)^2 (\nabla^{2} \phi) \\
& - (12+8\phi) (\nabla_{\alpha} \nabla_{\beta} \phi)
(\nabla^{\alpha} \nabla^{\beta} \phi)
+ (12+8\phi) (\nabla^{2} \phi)^2 \\
& + (24+16\phi) (\nabla^{\beta} \nabla^{\alpha} \phi)
(\nabla_{\alpha} \phi) (\nabla_{\beta} \phi)
- 4 (1+\phi) R (\nabla^{2} \phi) + 6 (\nabla \phi)^4 \\
& + (8+8\phi)\!\left[T_{\alpha}(\nabla^{\alpha}\phi)(\nabla^2\phi)
- T_{\beta}(\nabla_{\alpha}\nabla^{\beta}\phi)\nabla^{\alpha}\phi
- T_{\mu\alpha\beta}\nabla^{\alpha}\phi (\nabla^{\mu}\nabla^{\beta}\phi)\right] \\
& + (12+8\phi)\!\left[T_{\alpha}\nabla^{\alpha}\phi (\nabla\phi)^2
+ T_{\alpha\beta\mu}\nabla^{\alpha}\phi(\nabla^{\mu}\nabla^{\beta}\phi)\right] \\
& + \phi \Big[ 8T^{\beta}{}_{\alpha}{}^{\mu}R_{\mu\beta}\nabla^{\alpha}\phi
+ 8T^{\mu}R_{\mu\alpha}\nabla^{\alpha}\phi
- 4T^{\beta\mu\nu}R_{\mu\nu\alpha\beta}\nabla^{\alpha}\phi
- 4T_{\alpha}R\nabla^{\alpha}\phi \Big] \Big\} \,.
\end{aligned}
\end{equation}
The full action for the regularized {ECGB} (rECGB) theory is then constructed as 
\begin{equation}
\label{fullaction2}
  L_{\text{rECGB}} = \alpha_{0} \sqrt{-g} + \alpha_{1} \sqrt{-g}\,R
  + \alpha_2\,\left.\frac{d}{dn}\bigl(\sqrt{-\tilde{g}}\,\tilde{\mathcal{G}}\bigr)\right|_{n=4} \,.
\end{equation}
In the torsionless (Levi--Civita) sector, $L_{\text{rECGB}}$ reduces---up to a total-derivative boundary term---to the conformally regularized 4D Einstein--Gauss--Bonnet dynamics of Ref.~\cite{fernandes2020derivation}, reproducing Eq.~(27) therein.

\section{Equations of motion}\label{sec:field_eq}

The regularized GB action in four dimensions, Eq.~(\ref{fullaction2}), is extremely complex, which makes a direct derivation of the equations of motion via variation particularly challenging. To gain insight into this problem, we first examine the simpler cases of two-dimensional regularized (torsion-free) Einstein gravity and EC gravity. The lessons learned from these lower-dimensional examples will provide valuable guidance for tackling the four-dimensional GB case.

\subsection{Regularized Einstein gravity}

We begin with regularized Einstein gravity in two dimensions. Under a conformal transformation, the Ricci scalar transforms as
\begin{equation}
    \sqrt{-g} \mathring{R} \to \sqrt{-\tilde{g}} \tilde{\mathring{R}}
    =\sqrt{-g}\exp({(n-2)\phi})
\Big(\mathring{R}-2(n-1)\mathring{\nabla}{}_{\alpha}\mathring{\nabla}{}^{\alpha}\phi-(n-1)(n-2)\mathring{\nabla}{}_{\alpha}\phi\mathring{\nabla}{}^{\alpha}\phi\Big)\,, 
\end{equation}
Taking the dimensional derivative, {it results in}
\begin{equation}
\begin{aligned}
 \frac{d  ( \sqrt{-\tilde{g}} \tilde{\mathring{R}}) }{dn}\bigg|_{n=2}
 =&\sqrt{-g} \Big\{\phi\left(\mathring{R}-2\mathring{\nabla}{}_{\alpha}\mathring{\nabla}{}^{\alpha}\phi\right)-2\mathring{\nabla}{}_{\alpha}\mathring{\nabla}{}^{\alpha}\phi-\mathring{\nabla}{}_{\alpha}\phi\mathring{\nabla}{}^{\alpha}\phi\Big\} \\[0.3em]
 =& \sqrt{-g}\left(\phi \mathring{R}+\mathring{\nabla}{}_{\alpha}\phi\mathring{\nabla}{}^{\alpha}\phi\right)+ \text{boundary term} \,.
\end{aligned}
\end{equation}
The regularized Einstein--Hilbert action is then given by 
\begin{equation}
 S=\int d^2x  \sqrt{-g}\left(\phi \mathring{R}+\mathring{\nabla}{}_{\alpha} \phi \mathring{\nabla}{}^{\alpha} \phi\right) \,,
\end{equation}
which leads to the equations of motion obtained by varying with respect to $\phi$ and $g_{\mu\nu}$, respectively,
\begin{equation}
\label{2d_einstein_scalar}
\mathring{R}-2\, \mathring{\square} \phi=0 \,, 
\end{equation}
and 
\begin{equation}
\label{2d_einstein_metric}
\mathring{\nabla}{}_{\mu} \phi \mathring{\nabla}{}_{\nu} \phi-\mathring{\nabla}{}_{\mu} \mathring{\nabla}{}_{\nu} \phi+g_{\mu \nu}\left(\mathring{\square} \phi-\frac{1}{2}(\mathring{\nabla} \phi)^2\right)=0 \,.
\end{equation}

These equations of motion can also be derived through an alternative approach that exploits the conformal properties of the Einstein tensor:
\begin{align}
\delta \tilde{g}^{\mu\nu}(g^{\mu\nu},\phi) 
= \exp(-2\phi)\,\delta g^{\mu\nu} -2\,\tilde{g}^{\mu\nu}\,\delta\phi.
\end{align}
First, by varying the action with respect to the scalar field $\phi$, we get
\begin{equation}
\begin{aligned}
  \delta_{\phi}  \sqrt{-\tilde{g}} \tilde{\mathring{R}}
  =&\sqrt{-\tilde{g}}\, \tilde{\mathring{G}}{}_{\mu\nu}\,\delta_{\phi} \tilde{g}^{\mu\nu}
  =\sqrt{-\tilde{g}}\, \tilde{\mathring{G}}{}_{\mu\nu}\,
   \delta_{\phi} (\exp(-2\phi){g}^{\mu\nu})\\
  =&-2\sqrt{-\tilde{g}}\, \tilde{\mathring{G}}{}_{\mu\nu}\tilde{g}^{\mu\nu}\delta\phi
  =(n-2)\sqrt{-\tilde{g}} \tilde{\mathring{R}}\, \delta\phi \,.   
\end{aligned}
\end{equation}
{Then, we differentiate the equation} with respect to $n$, and evaluate it at $n=2$, leading to the scalar field equation:
\begin{equation}
  \tilde{\mathring{R}}=0  \quad\Rightarrow\quad  \mathring{R}-2\, \mathring{\square} \phi=0 \,.
\end{equation}
This coincides with Eq.~\eqref{2d_einstein_scalar} obtained by direct variation.
Next, {we} vary the action with respect to the metric $g_{\mu\nu}$, {which gives}
\begin{equation}
\begin{aligned}
  \delta_{g}  \sqrt{-\tilde{g}} \tilde{\mathring{R}}
  &= \sqrt{-\tilde{g}}\, \tilde{\mathring{G}}{}_{\mu\nu}\,\delta \tilde{g}^{\mu\nu}
  = \sqrt{-\tilde{g}}\, \tilde{\mathring{G}}{}_{\mu\nu}\,
    \delta_{g} (\exp(-2\phi){g}^{\mu\nu})
  = \sqrt{-\tilde{g}}\, \tilde{\mathring{G}}{}_{\mu\nu}\exp(-2\phi)\delta{g}^{\mu\nu}\\[0.3em]
  &=\sqrt{-g}\, \exp((n-2)\phi)\tilde{\mathring{G}}{}_{\mu\nu}\delta {g}^{\mu\nu} \,,
\end{aligned}
\end{equation}
where the conformally transformed Einstein tensor $\tilde{\mathring{G}}{}_{\mu\nu}$ is given by
\begin{equation}
   \tilde{\mathring{G}}{}_{\mu\nu}= \mathring{G}{}_{\mu\nu}+g_{\mu\nu}\left((n-2)\mathring{\nabla}{}^2\phi+\frac{(n-2)(n-3)}{2}(\mathring{\nabla} \phi)^2\right)+(n-2)\left(\mathring{\nabla}{}_{\mu}\phi\mathring{\nabla}{}_{\nu}\phi-\mathring{\nabla}{}_{\mu}\mathring{\nabla}{}_{\nu}\phi\right) \,.
\end{equation}
{By the similar procedure to differentiate}  {$\tilde{\mathring{G}}{}_{\mu\nu}$} with respect to $n$, and then setting $n=2$, {we would obtain}  the metric field equation
\begin{equation}
 \sqrt{-g}\,\frac{d \tilde{\mathring{G}}{}_{\mu\nu} }{dn}\bigg|_{n=2}
 =\sqrt{-g} \Biggl[\mathring{\nabla}{}_{\mu} \phi \mathring{\nabla}{}_{\nu} \phi-\mathring{\nabla}{}_{\mu} \mathring{\nabla}{}_{\nu} \phi+g_{\mu \nu}\left(\mathring{\square} \phi-\frac{1}{2}(\mathring{\nabla} \phi)^2\right)\Biggr]=0 \,.
\end{equation}
{Consequently,} this expression {also} coincides with Eq.~\eqref{2d_einstein_metric} obtained by direct variation, confirming the consistency of the approach.

\subsection{Regularized Einstein--Cartan gravity}
\label{sec:EC-regularized}

We now extend our analysis to include torsion by considering  EC gravity in two dimensions. 
{The results of dimensional regularization and variations are given in this section, 
the intermediate steps of calculation can be found in Appendix \ref{app:dim-reg-details}.}
Under a conformal transformation, the Ricci scalar with torsion transforms as
\begin{equation}
\begin{aligned}
\sqrt{-g}\, R 
\to \sqrt{-\tilde{g}}\, \tilde{R}
    &= \sqrt{-g}\,\exp\bigl((n-2)\phi\bigr)\times \\
    &\quad\times \Big(R-2(n-1)\nabla_{\alpha}\nabla^{\alpha}\phi 
             - (n-1)(n-2)\nabla_{\alpha}\phi\nabla^{\alpha}\phi-2T^{\alpha}\nabla_{\alpha}\phi\Big)\,. 
\end{aligned}
\end{equation}
The dimensional regularization  yields
\begin{equation}
\label{eq:dRdn-2d-EC}
\begin{aligned}
 \frac{d  \bigl( \sqrt{-\tilde{g}}\, \tilde{R}\bigr) }{dn}\bigg|_{n=2}
=&\;\sqrt{-g}\left\{\phi\bigl( \mathring{R}+T\bigr)+(\mathring\nabla \phi)^2\right\}+\text{boundary term}\,,
\end{aligned}
\end{equation}
Here, $T$ is the torsion scalar, defined by 
\begin{equation}
\label{def:torsion-scalar}
  T \;\equiv\;
  \frac{1}{4}\,T_{\alpha\beta\gamma}T^{\alpha\beta\gamma}
  +\frac{1}{2}\,T_{\alpha\beta\gamma}T^{\gamma\beta\alpha}
  - T_\alpha T^\alpha \,,
\end{equation}
where $T^{\alpha}$ is the torsion vector $T^{\alpha}\equiv T^{\mu\alpha}{}_{\mu}$.
Hence, the regularized action is 
\begin{equation}
\label{regularecgravity}
 S_{\text{rEC}}=\int d^2x\, \sqrt{-g}\left\{\phi\bigl( \mathring{R}+T\bigr)+(\mathring\nabla \phi)^2\right\}\,.
\end{equation}
Varying this action with respect to $\phi$, $g_{\mu\nu}$, and $K^{\mu}{}_{\nu\rho}$ yields the equations of motion:
\begin{equation}
\label{2d_ec_scalar}
 \mathring{R}+T-2\mathring\nabla{}^2 \phi=0\,,
\end{equation}
\begin{equation}\label{2d_ec_metric}
\begin{aligned}
&
  \frac{1}{2}\left( -g_{\mu\nu}\phi T-K_{\mu}{}^{\alpha\beta}K_{\alpha\beta\nu}\phi-K_{\nu}{}^{\alpha\beta}K_{\alpha\beta\mu}\phi+K_{\alpha}{}^{\beta}{}_{\beta}K_{\mu}{}^{\alpha}{}_\nu\phi+K_{\alpha}{}^{\beta}{}_{\beta}K_{\nu}{}^{\alpha}{}_\mu\phi\right) \\
  &\quad+\mathring\nabla_\mu \phi \mathring\nabla_\nu \phi-\mathring\nabla_\mu \mathring\nabla_\nu \phi+g_{\mu \nu}\left(\mathring\square \phi-\frac{1}{2}(\mathring\nabla \phi)^2\right)=0\,,
\end{aligned}
\end{equation}
\begin{equation}
\label{2d_ec_torsion}
   \delta_{\nu}^{\rho} K_{\mu}+K_{\mu}{}^{\rho}{}_{\nu}-\delta_{\mu}^{\rho}K_{\nu}-K_{\nu}{}^{\rho}{}_{\mu}=0\,,
\end{equation}
respectively.
These equations of motion can also be obtained by exploiting the conformal properties of the curvature tensors. The variation of the conformally transformed Ricci scalar is given by
\begin{equation}
\begin{aligned}
   \delta \bigl(\sqrt{-\tilde{g}}\, \tilde{R}\bigr)
   =&\;\sqrt{-\tilde{g}} \Biggl[\tilde{\mathring{G}}_{\mu\nu}+\frac{1}{2} \biggl(-\tilde{g}_{\mu\nu}\tilde{K}_{\alpha\gamma\beta}\tilde{K}^{\alpha\beta\gamma}-\tilde{g}_{\mu\nu}\tilde{K}^{\alpha\beta}{}_{\alpha}\tilde{K}_{\beta}{}^{\gamma}{}_{\gamma}-\tilde{K}_{\alpha\beta\nu}\tilde{K}_{\mu}{}^{\alpha\beta}\\
   &\qquad\qquad +\tilde{K}_{\alpha}{}^{\beta}{}_{\beta}\tilde{K}_{\mu}{}^{\alpha}{}_{\nu}-\tilde{K}_{\alpha\beta\mu}\tilde{K}_{\nu}{}^{\alpha\beta}
   +\tilde{K}_{\alpha}{}^{\beta}{}_{\beta}\tilde{K}_{\nu}{}^{\alpha}{}_{\mu}\biggr)\Biggr] \delta \tilde{g}^{\mu\nu}\\
   &\; +\sqrt{-\tilde{g}}\left(\tilde{g}^{\nu\rho}\tilde{K}_{\mu}+\tilde{K}_{\mu}{}^{\rho\nu}-\delta_{\mu}^{\rho}\tilde{K}^{\nu}-\tilde{K}^{\nu\rho}{}_{\mu}\right)\delta\tilde{K}^{\mu}{}_{\nu\rho}\,.
\end{aligned}
\end{equation}

Varying the action with respect to the scalar field $\phi$ leads to
\begin{equation}
\label{eq:delta-phi-metric-affine-compact}
\begin{aligned}
   \delta_{\phi}  \bigl(\sqrt{-\tilde{g}}\, \tilde{R}\bigr)
   =&\;(n-2)\bigl(\tilde{T}+\tilde{\mathring{R}}\bigr)\sqrt{-\tilde{g}}\,\delta\phi\;.
\end{aligned}
\end{equation}
By taking the derivative with respect to $n$ and then evaluating at $n=2$, we obtain 
\begin{equation}
 \tilde{T}+\tilde{\mathring{R}}=0\,.
\end{equation}
Considering that the torsion scalar vanishes identically in two dimensions, which we have proved in Appendix \ref{Appendix:vanishing_of_the_torsion_scalar_in_two_dimensions}, 
one finds that the scalar equation derived here is equivalent to \eqref{2d_ec_scalar}.

For metric $g_{\mu\nu}$, the variational form of the Lagrangian density reads
\begin{equation}
\begin{aligned}
   \delta_{g}  \bigl(\sqrt{-\tilde{g}}\, \tilde{R}\bigr)
   =&\;\sqrt{-\tilde{g}} \Biggl[\tilde{\mathring{G}}_{\mu\nu}
   +\frac{1}{2} \biggl(
   -\tilde{g}_{\mu\nu}\tilde{K}_{\alpha\gamma\beta}\tilde{K}^{\alpha\beta\gamma}
   -\tilde{g}_{\mu\nu}\tilde{K}^{\alpha\beta}{}_{\alpha}\tilde{K}_{\beta}{}^{\gamma}{}_{\gamma} 
   -\tilde{K}_{\alpha\beta\nu}\tilde{K}_{\mu}{}^{\alpha\beta}\\
   &\;+\tilde{K}_{\alpha}{}^{\beta}{}_{\beta}\tilde{K}_{\mu}{}^{\alpha}{}_{\nu}
   -\tilde{K}_{\alpha\beta\mu}\tilde{K}_{\nu}{}^{\alpha\beta} 
   +\tilde{K}_{\alpha}{}^{\beta}{}_{\beta}\tilde{K}_{\nu}{}^{\alpha}{}_{\mu}\biggr)\Biggr] \exp(-2\phi)\,\delta g^{\mu\nu}\,.
\end{aligned}
\end{equation}
Computing the derivative and setting $n=2$ yields
\begin{equation}
\begin{aligned}
   & \sqrt{-g} \Biggl[ \frac{1}{2}\left( -g_{\mu\nu}\phi T-K_{\mu}{}^{\alpha\beta}K_{\alpha\beta\nu}\phi-K_{\nu}{}^{\alpha\beta}K_{\alpha\beta\mu}\phi+K_{\alpha}{}^{\beta}{}_{\beta}K_{\mu}{}^{\alpha}{}_\nu\phi+K_{\alpha}{}^{\beta}{}_{\beta}K_{\nu}{}^{\alpha}{}_\mu\phi\right)\Biggr] \\[0.2em]
   &\quad+\sqrt{-g} \Biggl[\mathring\nabla_\mu \phi \mathring\nabla_\nu \phi-\mathring\nabla_\mu \mathring\nabla_\nu \phi+g_{\mu \nu}\left(\mathring\square \phi-\frac{1}{2}(\mathring\nabla \phi)^2\right)\Biggr]=0\,.
\end{aligned}
\end{equation}
This result exactly matches the metric field equation \eqref{2d_ec_metric} obtained through the direct variation, confirming the consistency of our approach.

It is important to note that in two dimensions, the torsion-dependent terms vanish identically, i.e.,
\begin{equation}
 \frac{1}{2}\left( -g_{\mu\nu}\phi T-K_{\mu}{}^{\alpha\beta}K_{\alpha\beta\nu}\phi-K_{\nu}{}^{\alpha\beta}K_{\alpha\beta\mu}\phi+K_{\alpha}{}^{\beta}{}_{\beta}K_{\mu}{}^{\alpha}{}_\nu\phi+K_{\alpha}{}^{\beta}{}_{\beta}K_{\nu}{}^{\alpha}{}_\mu\phi\right)\equiv 0\,.
\end{equation}
Therefore, the metric field equation can be simplified to:
\begin{equation}
 \sqrt{-g} \Biggl[\mathring\nabla_\mu \phi \mathring\nabla_\nu \phi-\mathring\nabla_\mu \mathring\nabla_\nu \phi+g_{\mu \nu}\left(\mathring\square \phi-\frac{1}{2}(\mathring\nabla \phi)^2\right)\Biggr]=0\,.
\end{equation}
This result is identical to the metric field equation \eqref{2d_einstein_metric} obtained in the  Einstein gravity case. This equivalence arises because in two dimensions, the torsion scalar $T$
identically vanishes, reducing the EC action (\ref{regularecgravity})
to the regularized  Einstein--Hilbert action
\begin{equation}
 S_{\text{rEH}}=\int d^2x\, \sqrt{-g}\left(\phi\, \mathring{R}+(\mathring\nabla \phi)^2\right)\,,
\end{equation}
thus yielding identical field equations for the metric.

For the contortion tensor $K^{\mu}{}_{\nu\rho}$, the equation of motion can be derived as follows:
\begin{equation}
\begin{aligned}
   \delta_{K^{\mu}{}_{\nu\rho}} \bigl(\sqrt{-\tilde{g}}\, \tilde{R}\bigr)
   =&\;\sqrt{-\tilde{g}}\left(\tilde{g}^{\nu\rho}\tilde{K}_{\mu}+\tilde{K}_{\mu}{}^{\rho\nu}-\delta_{\mu}^{\rho}\tilde{K}^{\nu}-\tilde{K}^{\nu\rho}{}_{\mu}\right)\delta\tilde{K}^{\mu}{}_{\nu\rho}\\
   =&\;\exp\!\bigl((n-2)\phi\bigr)\sqrt{-g}\left({g}^{\nu\rho}{K}_{\mu}+{K}_{\mu}{}^{\rho\nu}-\delta_{\mu}^{\rho}{K}^{\nu}-{K}^{\nu\rho}{}_{\mu}\right)\delta{K}^{\mu}{}_{\nu\rho}\,.
\end{aligned}
\end{equation}
Computing the derivative with respect to $n$ and setting $n = 2$ gives
\begin{equation}
\label{regulareccontorsion}
{g}^{\nu\rho}{K}_{\mu}+{K}_{\mu}{}^{\rho\nu}-\delta_{\mu}^{\rho}{K}^{\nu}-{K}^{\nu\rho}{}_{\mu}=0\,.
\end{equation}
This reproduces the torsion equation \eqref{2d_ec_torsion} obtained by direct variation. In two dimensions the expression vanishes identically, consistent with $T\equiv0$ and the reduction to the regularized Einstein--Hilbert action.

In the EC theory, an alternative approach is to take the tetrad $e^{a}{}_{\mu}$ and the spin connection $\omega^{a}{}_{b\mu}$ as the fundamental variables for variation. This formulation provides a more fundamental description of the gravitational field in the presence of torsion. The variation of the conformally transformed Lagrangian density with respect to these fundamental fields takes the compact form:
\begin{equation}
\delta \bigl(\sqrt{-\tilde{g}}\, \tilde{R}\bigr)
=\sqrt{-\tilde{g}}\tilde{U}^{\mu}{}_{a}\,\delta\tilde{e}^{a}{}_{\mu}
   +\sqrt{-\tilde{g}}\tilde{V}^{\mu}{}_{ab}\,\delta \tilde{\omega}^{ab}{}_{\mu}\,,
\end{equation}
where $\tilde{U}^{\mu}{}_{a}$ and $\tilde{V}^{\mu}{}_{ab}$ denote the coefficients arising from this variation:
\begin{equation}
\tilde{U}^{\mu}{}_{a} \equiv \frac{1}{2}\delta_{\mu_1\mu_2\alpha}^{\nu_1\nu_2\mu}\,\tilde{R}^{\mu_1\mu_2}{}_{\nu_1\nu_2}\,\tilde{e}_{a}{}^{\alpha}=  -2\tilde{R}^{\mu }{}_{a}+\tilde{R}\,\tilde{e}_{a}{}^{\mu}\,,
\end{equation}
\begin{equation}
\tilde{V}^{\mu}{}_{ab} \equiv \frac{1}{2}\delta_{\alpha\beta\mu_3}^{\mu\nu_2\nu_3}\,\tilde{T}^{\mu_3}{}_{\nu_2\nu_3}\,\tilde{e}_{a}{}^{\alpha}\tilde{e}_{b}{}^{\beta}= \tilde{T}^{\mu}{}_{ ab}+\tilde{T}_{b}\,\tilde{e}_{a}{}^{\mu}-\tilde{T}_{a}\,\tilde{e}_{b}{}^{\mu}\,.
\end{equation}
The variations of the conformally transformed fundamental fields are
\begin{equation}
  \delta  \tilde{e}^{a}{}_{\mu}
  =\delta \bigl( \exp(\phi)\,e^{a}{}_{\mu}\bigr)
  = \ \tilde{e}^{a}{}_{\mu}  \,\delta \phi
    +\exp(\phi)\,\delta e^{a}{}_{\mu}\,,
\end{equation}
and
\begin{equation}
\begin{aligned}
 \delta \tilde{\omega}^{ab}{}_{\mu}
 &=\delta {\omega}^{ab}{}_{\mu}+\delta  \left(e^a{}_{\mu}e^{b\nu}\partial_{\nu}\phi -e^b{}_{\mu}e^{a\nu}\partial_{\nu}\phi \right) \\[0.2em]
 &= \delta {\omega}^{ab}{}_{\mu}
 +e^{b\nu}\partial_{\nu}\phi\,  \delta  e^{a}{}_{\mu}
 - e^{a}{}_{\mu}e_{d}{}^{\nu}e^{b\alpha}\partial_{\nu}\phi\,\delta e^{d}{}_{\alpha}
 +e^a{}_{\mu}e^{b\nu}\partial_{\nu}\delta \phi   
 -(a\leftrightarrow b)\,.
\end{aligned}
\end{equation}

Varying the action with respect to the scalar field $\phi$ in the tetrad--spin-connection formalism gives
\begin{equation}
\label{eq:delta-phi-tetrad-compact}
\begin{aligned}
       \delta_{\phi} \bigl(\sqrt{-\tilde{g}}\, \tilde{R}\bigr)
       =&\;\sqrt{-g}\,(n-2)\tilde{R}\,\exp(n\phi)\,\delta \phi
       -\sqrt{-g}\,\mathring{\nabla}_{\nu}\Bigl[\exp\!\bigl((n-2)\phi\bigr)\,(2n-4)\,T^{\nu}\Bigr]\delta \phi\; \\ &+\text{boundary terms}\;.
\end{aligned}
\end{equation}
Taking the derivative with respect to $n$ and evaluating at $n=2$, the equation of motion of the scalar field would be
\begin{equation}
\tilde{R}\,\exp(2\phi)-2\,\mathring{\nabla}_{\nu}T^{\nu}=0,
\quad\text{{i.e.,}}\quad
\mathring{R}+T-2\mathring\nabla{}^2 \phi=0\,,
\end{equation}
{with the definition of $\mathring\nabla{}^2 := \mathring\nabla^{\mu}\mathring\nabla_{\mu}$,}
which is equivalent to Equation \eqref{2d_ec_scalar}.
Next, varying the action with respect to the tetrad field $e^{a}{}_{\mu}$ gives
\begin{equation}
\label{eq:delta-e-tetrad-compact}
\begin{aligned}
      & \delta_{e} \bigl(\sqrt{-\tilde{g}}\, \tilde{R}\bigr)\\
       =&\;\sqrt{-g} \Bigl[-2\tilde{R}^{\mu }{}_{\nu}+\tilde{R}\,\delta_{\nu}^{\mu}\Bigr]e_{a}{}^{\nu}\exp(n\phi)\,\delta e^{a}{}_{\mu}\\
       &\;+\sqrt{-g}\,\exp\!\bigl((n-2)\phi\bigr)\Bigl[2T^{\mu}{}_{a\beta} \mathring{\nabla}{}^{\beta}\phi+2e_{a}{}^{\mu}T_\beta \mathring{\nabla}{}^{\beta}\phi-2T_{a}\mathring{\nabla}{}^{\mu}\phi-2(n-2)T^{\mu}\mathring{\nabla}{}_{a}\phi\Bigr]\delta e^{a}{}_{\mu}\;.
\end{aligned} 
\end{equation}
Taking the derivative with respect to $n$ at $n=2$ yields an expression that can be organized as
\begin{equation}
\begin{aligned}
   &\frac{d}{dn}\Bigl(  \tilde{U}^{\mu}{}_{\alpha} \Bigr)\Big|_{n=2} {e}_{a}{}^{\alpha}\,  \exp(2\phi)
   +2\,\frac{d}{dn}\Bigl(\tilde{V}^{\mu}{}_{\alpha}{}^{\beta}-\tilde{V}^{\rho}{}_{\rho}{}^{\mu}\delta_{\alpha}^{\beta}\Bigr))\bigg|_{n=2}\mathring{\nabla}_{\beta}\phi\, e_{a}{}^{\alpha}\, \exp(2\phi)=0\\[0.3em]
   &\to\;
  \frac{d}{dn}\Bigl(  \tfrac{1}{2}\delta_{\mu_1\mu_2\alpha}^{\nu_1\nu_2\mu}\,\tilde{R}^{\mu_1\mu_2}{}_{\nu_1\nu_2}  \Bigr)\bigg|_{n=2}{e}_{a}{}^{\alpha}\,  \exp(2\phi)
  +2\,\frac{d}{dn}\Bigl(\tilde{V}^{\mu}{}_{\alpha}{}^{\beta}-\tilde{V}^{\rho}{}_{\rho}{}^{\mu}\delta_{\alpha}^{\beta}\Bigr)\bigg|_{n=2}\mathring{\nabla}_{\beta}\phi\, e_{a}{}^{\alpha}\, \exp(2\phi)=0\\[0.3em]
  &\to\; \Biggl[
  \frac{d}{dn}\Bigl(  \tfrac{1}{2}\delta_{\mu_1\mu_2\alpha}^{\nu_1\nu_2\mu}\,\tilde{R}^{\mu_1\mu_2}{}_{\nu_1\nu_2}  \Bigr)\bigg|_{n=2}
  +2\,\frac{d}{dn}\Bigl(\tilde{V}^{\mu}{}_{\alpha}{}^{\beta}-\tilde{V}^{\rho}{}_{\rho}{}^{\mu}\delta_{\alpha}^{\beta}\Bigr)\bigg|_{n=2}\mathring{\nabla}_{\beta}\phi\Biggr] e_{a}{}^{\alpha}\, \exp(2\phi)=0\,.
\end{aligned}
\end{equation}
This can be simplified to the explicit form of
\begin{equation}
\label{2d_ec_metric_full}
\begin{aligned}
&\delta  ^{\mu}_{\nu}\Bigl(-2\, \mathring\square \phi+\tfrac{1}{2}(\mathring\nabla \phi)^2\Bigr)
+2\Bigl(-\mathring\nabla{}^\mu \phi \,\mathring\nabla_\nu \phi+\mathring\nabla{}^{\mu} \mathring\nabla_\nu \phi\Bigr) \\
&\quad\quad\quad\quad\quad+\Bigl(-2\delta_{\nu}^{\mu}K_{\beta}+2K^{\mu}{}_{\beta\nu}+2K^{\mu}g_{\beta\nu}\Bigr)\mathring{\nabla}^{\beta} \phi=0\,.
\end{aligned}
\end{equation}
However, it is crucial to note that in two dimensions, the torsion-dependent terms vanish identically, {that is}
\begin{equation}
\label{2d_torsion_identity}
\Bigl(-2\delta_{\nu}^{\mu}K_{\beta}+2K^{\mu}{}_{\beta\nu}+2K^{\mu}g_{\beta\nu}\Bigr)\equiv 0\,.
\end{equation}
Consequently, Eq.~\eqref{2d_ec_metric_full} is reduced to the metric field equation \eqref{2d_ec_metric} obtained in the torsion-free case.

Finally, varying the action with respect to the spin connection $\omega^{a}{}_{b\mu}$ gives
\begin{equation}
\begin{aligned}
\delta_{\omega} \bigl(\sqrt{-\tilde{g}}\, \tilde{R}\bigr)
&=\sqrt{-\tilde{g}}\tilde{V}^{\mu}{}_{ab}\,\delta \tilde{\omega}^{ab}{}_{\mu}
 =\sqrt{-\tilde{g}}\tilde{V}^{\mu}{}_{ab}\,\delta {\omega}^{ab}{}_{\mu} \\
&=\sqrt{-g}\,\tilde{V}^{\mu}{}_{\alpha\beta}\, \exp\!\bigl((n-2)\phi\bigr)\,e_{a}^{\alpha}e_{b}^{\beta}\,\delta {\omega}^{ab}{}_{\mu}\,.
\end{aligned}
\end{equation}
Taking the derivative with respect to $n$ and setting $n=2$ yields, we find that the contribution vanishes due to
\begin{equation}
        \frac{d}{dn}\tilde{V}^{\mu}{}_{\alpha\beta}\equiv 0\,.
\end{equation}
This reflects our earlier observation that, in two dimensions, torsion-dependent terms vanish identically \eqref{2d_ec_torsion}.

\subsection{Regularized Einstein--Cartan--Gauss--Bonnet gravity}
Now we apply the technique of taking dimensional derivatives of the field equations obtained from variation with respect to the tetrad and spin connection to the regularized ECGB gravity.
The variation of the conformally transformed GB Lagrangian density decomposes into tetrad and spin connection contributions:
\begin{equation}
\begin{aligned}
   \delta\!\left(\sqrt{-\tilde{g}}\, \tilde{\mathcal{G}}\right)
   &= \sqrt{-\tilde{g}}\, \tilde{H}^{\mu}{}_{\alpha}\,\tilde{e}_{a}{}^{\alpha}\, \delta  \tilde{e}^{a}{}_{\mu}
   + \sqrt{-\tilde{g}}\,\tilde{F}^{\mu}{}_{ab}\,\delta \tilde{\omega}^{ab}{}_{\mu}\,.
\end{aligned}
\end{equation}
where the coefficients are defined {by}
\begin{equation}
\begin{aligned}
 \tilde{H}^{\mu}{}_{\alpha}
 &= \frac{1}{4}\,\delta_{\mu_1\mu_2\mu_3\mu_4\alpha}^{\nu_1\nu_2\nu_3\nu_4\mu}\,
 \tilde{R}^{\mu_1\mu_2}{}_{\nu_1\nu_2}\,
 \tilde{R}^{\mu_3\mu_4}{}_{\nu_3\nu_4} \\[0.2em]
 &=
 -2\Biggl[\,2\tilde{R}^{\mu\beta\rho\sigma} \tilde{R}_{\rho\sigma\alpha\beta}
   -4\tilde{R}^{\mu\rho}\tilde{R}_{\rho\alpha}
   -4\tilde{R}^{\mu\rho}{}_{\alpha}{}^{\sigma}\tilde{R}_{\sigma\rho}
   +2\tilde{R}^{\mu}{}_{\alpha}\tilde{R}
   -\frac{1}{2}\delta^{\mu}_{\alpha}\tilde{\mathcal{G}}\Biggr]\,,
\end{aligned}
\end{equation}
and
\begin{equation}
\begin{aligned}
 \tilde{F}^{\mu}{}_{ab}
 = \frac{1}{2}\,\delta_{\alpha\beta\mu_3\mu_4\mu_5}^{\mu\nu_2\nu_3\nu_4\nu_5}\,
 \tilde{R}^{\mu_3\mu_4}{}_{\nu_2\nu_3}\,
 \tilde{T}^{\mu_5}{}_{\nu_4\nu_5}\,\tilde{e}_{a}{}^{\alpha}\tilde{e}_{b}{}^{\beta}\,.
\end{aligned}
\end{equation}
The variations of the conformally transformed fundamental fields {of tetrad and spin connection are respectively} 
\begin{equation}
  \delta  \tilde{e}^{a}{}_{\mu}
  = \delta\bigl( \exp(\phi)\,e^{a}{}_{\mu}\bigr)
  = \exp(\phi)\,e^{a}{}_{\mu}\,\delta \phi+\exp(\phi)\,\delta e^{a}{}_{\mu}\,,
\end{equation}
and
\begin{equation}
\begin{aligned}
 \delta \tilde{\omega}^{ab}{}_{\mu}
 &= \delta {\omega}^{ab}{}_{\mu}
    + \delta \!\left( e^a{}_{\mu}e^{b\nu}\partial_{\nu}\phi
                     - e^b{}_{\mu}e^{a\nu}\partial_{\nu}\phi \right) \\[0.2em]
 &= \delta {\omega}^{ab}{}_{\mu}
    + e^{b\nu}\partial_{\nu}\phi\, \delta  e^{a}{}_{\mu}
    - e^{a}{}_{\mu}e^{\nu}{}_{d}e^{b\alpha}\partial_{\nu}\phi\,\delta e^{d}{}_{\alpha}
    + e^a{}_{\mu}e^{b\nu}\partial_{\nu}\delta \phi \; -(a\leftrightarrow b)\,.
\end{aligned}
\end{equation}
{Now the variation of GB term} with respect to the scalar field $\phi$ yields 
\begin{equation}
\begin{aligned}
 \delta_{\phi}\!\left(\sqrt{-\tilde{g}}\, \tilde{\mathcal{G}}\right)
 &= \sqrt{-\tilde{g}}\,(n-4)\Biggl[\delta_{\mu_1\mu_2\mu_3\mu_4}^{\nu_1\nu_2\nu_3\nu_4}\,\frac{1}{4}\,
    \tilde{R}^{\mu_1\mu_2}{}_{\nu_1\nu_2}\tilde{R}^{\mu_3\mu_4}{}_{\nu_3\nu_4}\Biggr]\delta \phi \\[0.2em]
 &\quad - 2\sqrt{-g}\,\mathring{\nabla}_{\nu}\!\Bigl(\exp(n\phi)\,\tilde{F}^{\mu}{}_{ab}\, e^a{}_{\mu}e^{b\nu}\Bigr) \\[0.2em]
 &= \sqrt{-\tilde{g}}\,(n-4)\Biggl[\delta_{\mu_1\mu_2\mu_3\mu_4}^{\nu_1\nu_2\nu_3\nu_4}\,\frac{1}{4}\,
    \tilde{R}^{\mu_1\mu_2}{}_{\nu_1\nu_2}\tilde{R}^{\mu_3\mu_4}{}_{\nu_3\nu_4}\Biggr]\delta \phi \\[0.2em]
 &\quad - \sqrt{-g}\,\mathring{\nabla}_{\nu}\!\Bigl((n-4)\exp(n\phi)\,
    \delta_{\beta\mu_3\mu_4\mu_5}^{\nu_2\nu_3\nu_4\nu_5}\,
    \tilde{R}^{\mu_3\mu_4}{}_{\nu_2\nu_3}\tilde{T}^{\mu_5}{}_{\nu_4\nu_5}\, \tilde{g}^{\beta\nu}\Bigr)\delta\phi\,.
\end{aligned}
\end{equation}
{By} taking the dimensional derivative and evaluating {the value} at $n=4$, {we obtain} the scalar field equation
\begin{equation}
\label{regular-ecg-phi}
 \sqrt{-\tilde{g}}\,\tilde{\mathcal{G}}
 \;-\; \sqrt{-g}\,\mathring{\nabla}_{\nu}\!\Bigl(\exp(4\phi)\,
 \delta_{\beta\mu_3\mu_4\mu_5}^{\nu_2\nu_3\nu_4\nu_5}\,
 \tilde{R}^{\mu_3\mu_4}{}_{\nu_2\nu_3}\tilde{T}^{\mu_5}{}_{\nu_4\nu_5}\, \tilde{g}^{\beta\nu}\Bigr)=0\,.
\end{equation}
In addition, for the tetrad variation, the dimensional derivative at $n=4$ yields
\begin{equation}
\begin{aligned}
 \frac{d}{dn}\!\left(\tilde{H}^{\mu}{}_{\alpha}\right) {e}_{a}{}^{\alpha}\, \exp(4\phi)
 \;+\; 2\,\frac{d}{dn}\!\Bigl(\tilde{F}^{\mu}{}_{\alpha}{}^{\beta}-\tilde{F}^{\rho}{}_{\rho}{}^{\mu}\delta_{\alpha}^{\beta}\Bigr)
 \mathring{\nabla}_{\beta}\phi \, e_{a}{}^\alpha \exp(4\phi)=0\,,
\end{aligned}
\end{equation}
{and} the variation with respect to the spin connection  {produces}
\begin{equation}
\begin{aligned}
\delta_{\omega}\!\left(\sqrt{-\tilde{g}}\, \tilde{\mathcal{G}}\right)
&= \sqrt{-\tilde{g}}\,\tilde{F}^{\mu}{}_{ab}\,\delta \tilde{\omega}^{ab}{}_{\mu}
 = \sqrt{-\tilde{g}}\,\tilde{F}^{\mu}{}_{ab}\,\delta {\omega}^{ab}{}_{\mu} \\
&= \sqrt{-g}\,\tilde{F}^{\mu}{}_{\alpha\beta}\,\exp\!\bigl((n-2)\phi\bigr)\,e_{a}{}^{\alpha}e_{b}{}^{\beta}\,\delta {\omega}^{ab}{}_{\mu}\,.
\end{aligned}
\end{equation}
{Again, by} computing the dimensional derivative and evaluating at $n=4$,  the spin connection equation {would be}
\begin{equation}
 \frac{d}{dn}\!\left(\tilde{F}^{\mu}{}_{\alpha\beta}\,\right)\bigg|_{n=4}\exp(2\phi)=0\,.
\end{equation}

\subsection{Full action and complete field equations}
Through the Lagrangian density \eqref{fullaction2} the complete regularized action is defined  by
\begin{equation}
S_{\text{rECGB}} = \int d^4x \Biggl[ \alpha_0 \sqrt{-g} + \alpha_1 \sqrt{-g} R + \alpha_2 \frac{d}{dn} \left( \sqrt{-\tilde{g}} \tilde{\mathcal{G}} \right) \bigg|_{n=4} \Biggr]\,.
\end{equation}
Accordingly, the equation of motion for the scalar field $\phi$ is the same as \eqref{regular-ecg-phi}
\begin{equation}\label{4d_ec_phi}
\sqrt{-\tilde{g}}\,\tilde{\mathcal{G}} \;-\; \sqrt{-g}\,\mathring{\nabla}_{\nu}\!\left(\exp(4\phi)\,\delta_{\beta\mu_3\mu_4\mu_5}^{\nu_2\nu_3\nu_4\nu_5}\tilde{R}^{\mu_3\mu_4}{}_{\nu_2\nu_3}\tilde{T}^{\mu_5}{}_{\nu_4\nu_5} \tilde{g}^{\beta\nu}\right)=0\,.
\end{equation}
Moreover, the variation with respect to the tetrad $e^a{}_\mu$ yields
\begin{equation}\label{tetrad equation}
\begin{aligned}
&\alpha_{0}\,e_{a}{}^{\mu}+\alpha_{1}\Bigl[-2{R}^{\mu }{}_{\nu}+{R}\delta_{\nu}^{\mu}\Bigr]e_{a}{}^{\nu}\\
&\quad +\alpha_{2}\Bigl[\tfrac{d}{dn}\!\left(\tilde{H}^{\mu}{}_{\alpha}\right)e_{a}{}^{\alpha}\exp(4\phi)
+2\,\tfrac{d}{dn}\!\left(\tilde{F}^{\mu}{}_{\alpha}{}^{\beta}-\tilde{F}^{\rho}{}_{\rho}{}^{\mu}\delta_{\alpha}{}^{\beta}\right)\mathring{\nabla}_{\beta}\phi\,e_{a}{}^{\alpha}\exp(4\phi)\Bigr]=0\,.
\end{aligned}
\end{equation}
In the coordinate basis, this becomes
\begin{equation}\label{4d_ec_metric}
\begin{aligned}
&\alpha_{0}\,\delta ^{\mu}_{\nu}+\alpha_{1}\Bigl[-2{R}^{\mu }{}_{\nu}+{R}\delta_{\nu}^{\mu}\Bigr] \\
&\quad +\alpha_{2}
        \Bigl[\tfrac{d}{dn}\!\left(\tilde{H}^{\mu}{}_{\nu}\right)\exp(4\phi)
               +2\,\tfrac{d}{dn}\!\left(\tilde{F}^{\mu}{}_{\nu}{}^{\beta}-\tilde{F}^{\rho}{}_{\rho}{}^{\mu}\delta_{\nu}^{\beta}\right)\mathring{\nabla}_{\beta}\phi\,\exp(4\phi)\Bigr]=0\,.
\end{aligned}
\end{equation}
Furthermore, variation with respect to the spin connection gives
\begin{equation}
\begin{aligned}
\delta_{\omega}L_{\text{rECGB}} 
&=\alpha_{1}\sqrt{-g}\,
  \Bigl[{T}^{\mu}{}_{ \alpha\beta} 
        + {T}_{\beta}{\delta}_{\alpha}^{\mu} 
        - {T}_{\alpha}{\delta}_{\beta}^{\mu}\Bigr]
  e_{a}{}^{\alpha}e_{b}{}^{\beta}\delta {\omega}^{ab}{}_{\mu} \\
& \quad\qquad +\alpha_{2}\sqrt{-g}\,\tilde{F}^{\mu}{}_{\alpha\beta}\exp((n-2)\phi)\,
        e_{a}{}^{\alpha}e_{b}{}^{\beta}\delta {\omega}^{ab}{}_{\mu}\,,
\end{aligned}
\end{equation}
and, finally, taking the dimensional derivative at $n=4$ yields the complete spin connection equation
\begin{equation}\label{4d_ec_spin}
\alpha_{1}\Bigl[{T}^{\mu}{}_{ \alpha\beta}+{T}_{\beta}{\delta}_{\alpha}^{\mu}-{T}_{\alpha}{\delta}_{\beta}^{\mu}\Bigr]
+\alpha_{2}\,\frac{d}{dn}\tilde{F}^{\mu}{}_{\alpha\beta}\,\exp(2\phi)=0\,.
\end{equation}
Together, Eqs.~\eqref{4d_ec_phi}, \eqref{4d_ec_metric}, and \eqref{4d_ec_spin} constitute the complete set of field equations for the rECGB theory in four dimensions. An explicit coordinate expression for the scalar equation, obtained by expanding \eqref{4d_ec_phi}, is given in Eq.~\eqref{tensor_equation_final} in Appendix \ref{Appendix:equations_of_motion}. 
At first sight this expression contains three terms with third covariant derivatives of the scalar, schematically
\[
 -8T_{\alpha}\nabla_{\beta}\nabla^{\beta}\nabla^{\alpha}\phi  \;,
 \quad
  8T_{\beta}\nabla^{\beta}\nabla^2\phi 
  \quad\text{and}\quad
  8T_{\beta\alpha\gamma}\nabla^{\gamma}\nabla^{\beta}\nabla^{\alpha}\phi.
\]
Using 
\[
[\nabla_{\mu},\nabla_{\nu}]\phi=-T^{\rho}{}_{\mu\nu}\nabla_{\rho}\phi
\quad\text{and}\quad
[\nabla_{\mu},\nabla_{\nu}]V^{\rho}=R^{\rho}{}_{\sigma\mu\nu}V^{\sigma}-T^{\lambda}{}_{\mu\nu}\nabla_{\lambda}V^{\rho},
\]
The sum of these three contributions can be rewritten entirely in terms of curvature and torsion tensors multiplying at most first and second derivatives of $\phi$, which can be found in Appendix \ref{Appendix:equations_of_motion}.   Since the tetrad and spin-connection equations \eqref{4d_ec_metric} and \eqref{4d_ec_spin} are manifestly second order, we conclude that all field equations of the rECGB theory are strictly second order in covariant derivatives of the dynamical fields. This guarantees the absence of Ostrogradsky-type instabilities associated with higher-than-second time derivatives in this model.

\subsection{Comparison with the coupling-constant rescaling method}
An alternative approach to regularizing the GB term, 
 inspired by Glavan and Lin's dimensional-regularization method, considers the $n$-dimensional action {of} 
\begin{equation}
S = \lim_{n\to 4} \int d^n x \sqrt{-g} \left( \alpha_0 + \alpha_1 R - \frac{\alpha_2}{n-4} \mathcal{G} \right)\,.
\end{equation}
Doing the variation with respect to the tetrad would give 
\begin{equation}
\begin{aligned}
\sqrt{-g}\alpha_{0}e_{a}{}^{\mu}\delta  {e}^{a}{}_{\mu}       
+\sqrt{-g} \alpha_{1}\bigg[-2{R}^{\mu }{}_{a}+{R}{e}_{a}{}^{\mu}\bigg] \delta  {e}^{a}{}_{\mu} 
- \sqrt{-g} \frac{\alpha_{2}}{(n-4)} {H}^{\mu}{}_{\alpha}{e}_{a}{}^{\alpha} \delta  {e}^{a}{}_{\mu}=0\,,
\end{aligned}
\end{equation}
which leads to the field equation
\begin{equation}
\begin{aligned}
\lim_{n\to4}\sqrt{-g}\alpha_{0}e_{a}{}^{\mu}    
+\sqrt{-g} \alpha_{1}\bigg[-2{R}^{\mu }{}_{a}+{R}{e}_{a}{}^{\mu}\bigg] 
- \sqrt{-g} \frac{\alpha_{2}}{(n-4)} {H}^{\mu}{}_{\alpha}{e}_{a}^{\alpha} =0\,,
\end{aligned}
\end{equation}
In addition, we can also obtain the trace of the field equation, that should be 
\begin{equation}
  \label{glavantrace}  4\alpha_{0}+2\alpha_{1}R-\alpha_{2}\mathcal{G}=0\,.
\end{equation}
Within our conformal regularization scheme, a corresponding trace equation follows from the combination $\delta^{\nu}_{\mu}\times\eqref{4d_ec_metric}-\alpha_{2}\times\eqref{4d_ec_phi}$; its explicit form {is} Eq.~\eqref{regular-ecg-trace-eq} { which can be found} in Appendix \ref{Appendix:equations_of_motion}.
{The trace equation is a little bit complicated in the rECGB model, however, it is not hard to show that} this trace can be reduced to \eqref{glavantrace} only in the torsion-free limit.

\section{Spherically symmetric solutions}

We seek static solutions possessing $\mathrm{SO}(3)$ symmetry. In Schwarzschild-like coordinates $x^{\mu}=\{t,r,\theta,\varphi\}$, the most general metric satisfying these conditions is given by
\begin{equation}
    ds^2=-A(r)^2\,dt^2+B(r)^2\,dr^2+
    r^2\bigl(d\theta^2+\sin^2\!\theta\,d\varphi^2\bigr)\,,
\end{equation}
{and} the tetrad $\{e^{i}\}$ {can be} chosen in a simple diagonal form as
\begin{equation}
    e^{0}=A(r)\,dt\;, \qquad
    e^{1}=B(r)\,dr\;, \qquad
    e^{2}=r\,d\theta\;, \qquad
    e^{3}=r\sin\theta\, d\varphi\,.
\end{equation}
{It is known that,} the Killing vectors  {associated with} $\mathrm{SO}(3)$ symmetry are
\begin{equation}
\begin{aligned}
\xi_{(0)} &= -\partial_{t}\,,\\
\xi_{(1)} &= \sin{\varphi}\,\partial_{\theta} 
     + \cos{\varphi}\cot{\theta}\,\partial_{\varphi}\,,\\
\xi_{(2)} &= -\cos{\varphi}\,\partial_{\theta} 
     + \sin{\varphi}\cot{\theta}\,\partial_{\varphi}\,,\\
\xi_{(3)} &= -\partial_{\varphi}\,.
\end{aligned}
\end{equation}
In RC geometry, the symmetry of a manifold equipped with both metric and torsion  naturally {satisfy} the following set of conditions
\begin{equation}
\begin{aligned}
\mathcal{L}_\xi e^{a}{}_{\alpha} &=\lambda^{a}{}_{b}\,e^{b}{}_{\alpha}\,,\\[0.3em]
\mathcal{L}_\xi \omega^{a}{}_{b\mu} &=-D_{\mu} \lambda^{a}{}_{b}\,,\\[0.3em]
\end{aligned}
\end{equation}
or, equivalently, in the coordinate basis
\begin{equation}
\begin{aligned}
\mathcal{L}_\xi g_{\mu\nu}&=0\,,\\[0.3em]
\mathcal{L}_\xi \Gamma^{\rho}{}_{\mu\nu}&=0\,.
\end{aligned}
\end{equation}
The most general $\mathrm{SO}(3)$–invariant spin connection compatible with the above conditions can be parameterized by eight radial functions (parity-even/odd pairs)
\begin{equation}\label{Eq:parameterized spin connection}
\begin{matrix}
\omega^{a}{}_{b t}=
\begin{pmatrix}
0 & f_1 & 0 & 0 \\
f_1 & 0 & 0 & 0 \\
0 & 0 & 0 & f_2 \\
0 & 0 & -f_2 & 0
\end{pmatrix}\!,
&\qquad
\omega^{a}{}_{b r}=
\begin{pmatrix}
0 & g_1 & 0 & 0 \\
g_1 & 0 & 0 & 0 \\
0 & 0 & 0 & g_2 \\
0 & 0 & -g_2 & 0
\end{pmatrix}\!, \\[1.2cm]
\omega^{a}{}_{b \theta}=
\begin{pmatrix}
0 & 0 & p_1 & q_2 \\
0 & 0 & q_1 & -p_2 \\
p_1 & -q_1 & 0 & 0 \\
q_2 & p_2 & 0 & 0
\end{pmatrix}\!,
&\qquad
\omega^{a}{}_{b \varphi}=\sin\theta\,
\begin{pmatrix}
0 & 0 & -q_2 & p_1 \\
0 & 0 & p_2 & q_1 \\
-q_2 & -p_2 & 0 & -\cot \theta \\
p_1 & -q_1 & \cot \theta & 0
\end{pmatrix}\!;
\end{matrix}
\end{equation}
{see, e.g., Obukhov (App.~C, Eqs.~(117)–(119)) \cite{obukhov2023poincare} and the classification in McNutt–Ramírez–Iosifidis \cite{mcnutt2024symmetries}.
{So Eq.\eqref{Eq:parameterized spin connection}} introduces eight additional arbitrary functions of the radial coordinate,
\begin{equation}
f_1(r),\; f_2(r),\; g_1(r),\; g_2(r),\; p_1(r),\; p_2(r),\; q_1(r),\; q_2(r)\,,
\end{equation}
and the inclusion of these functions reflects the additional degrees of freedom associated with torsion in RC geometry, compared to the purely metric-based formulation of general relativity.

We now proceed to find exact black hole solutions within the {rECGB} framework. 
There are two standard routes to the spherical field equations: 
(i) vary the full four-dimensional equations  ~\eqref{4d_ec_phi}, \eqref{4d_ec_metric}, and~\eqref{4d_ec_spin} and then impose the ansatz; 
(ii) use the ansatz at the action level and vary the reduced Lagrangian. 
In what follows we adopt route (ii) because it is technically more efficient for constructing exact solutions; under the same symmetry assumptions, both routes are equivalent at the level of the ansatz.

Throughout this section we work in the parity-even and rotation-invariant sector by requiring invariance not only under the $\mathrm{SO}(3)$ isometries but also under spatial reflections. 
This upgrades the symmetry group to the full rotation group $\mathrm{O}(3)$ and, in the standard spherical decomposition of Poincaré-gauge connections, eliminates the parity-odd (axial) pieces. 
Concretely, we consistently set
\[
f_2(r)=g_2(r)=p_2(r)=q_2(r)=0 \,,
\]
so that only the even-parity components $\{f_1,g_1,p_1,q_1\}$ survive together with the metric/scalar functions $(A,B,\phi)$.
With these assumptions, we substitute the spherically symmetric ansatz directly into the action and vary the resulting one-dimensional functional. 
The reduced action reads
\begin{equation}
  \mathcal{S} \;=\; 4\pi \int dt\,dr\, L\big(A,B,\phi;f_1,g_1,p_1,q_1;r\big)\,,
\end{equation}
where the explicit Lagrangian $L$ is lengthy and collected in Appendix \ref{Appendix:reduced_lagrangian_for_the_spherical_ansatz}, see Eq.~\eqref{eq:Lreduced}.
Applying the Euler--Lagrange equations to $L$ yields seven ordinary differential equations for the unknown functions; their explicit forms are also listed in Appendix \ref{Appendix:reduced_lagrangian_for_the_spherical_ansatz}.

\subsection{Solution 1}

We adopt an ansatz for the spin connection inspired by the work of Chen et al.~\cite{chen2018new}, which studied spherically symmetric solutions in EC gravity. Our ansatz is
\begin{equation}
\begin{aligned}
    & f_1 = \frac{A'}{B},\quad
      f_2 = 0,\quad
      g_1 = -\frac{B}{A}G,\quad
      g_2 = 0,\,\\
    & p_1 = -\frac{r}{A}P,\quad
      p_2 = 0,\quad
      q_1 = -\frac{1}{B},\quad
      q_2 = 0 \,.
\end{aligned}
\end{equation}
The non-vanishing torsion components are
\begin{equation}
    T^{r}{}_{t r} = G,\quad
    T^{\theta}{}_{t \theta} = P,\quad
    T^{\varphi}{}_{t \varphi} = P \,.
\end{equation}
Substituting this ansatz into the reduced action and varying with respect to the metric functions and torsion components, we obtain the field equations. Under the special parameter relation $2\alpha_{0}\alpha_{2}+3\alpha_{1}^2=0$, we find the exact solution
\begin{equation}
\begin{aligned}
   A^2&=\frac{1}{B^2}=-\frac{\alpha_1\bigl(1+r^2 c_1\bigr)}{2\alpha_2 c_1}\,,\\[0.3em]
   \phi &= \frac{r\,\operatorname{arctanh}\!\sqrt{1+r^2 c_1}\,\sqrt{1/r^2+c_1}}
            {\sqrt{1+r^2 c_1}} + c_3 - \ln r \,,\\[0.3em]
   P &= \frac{1}{2}\,\sqrt{\frac{2 \alpha_1}{\alpha_2}
     +\frac{\alpha_1^2}{r^2 \alpha_2^2 c_1^2}
     +\frac{\alpha_1^2}{\alpha_2^2 c_1}
     +\frac{2 \alpha_1}{r^2 \alpha_2 c_1}
     +\frac{4 c_2 \sqrt{1+r^2 c_1}}{r}}\,,\\[0.3em]
   G&=0 \,.
\end{aligned}
\end{equation}
In the convention of Ref.~\cite{cvetkovic2018black}, which constructs a 5D black hole with torsion in the 5DECGB model in Lovelock gravity with the Lagrangian to be 
\begin{equation}
  L_{\text{5DECGB}} = \alpha_{0} \sqrt{-g} + \alpha_{1} \sqrt{-g}\,R
  + \alpha_2\,\sqrt{-{g}}\,{\mathcal{G}}.
\end{equation} 
The torsionful solution branch obeys
\begin{equation}
  \alpha_1^2 - 2\,\alpha_0\,\alpha_2 = 0 \,.
\end{equation}
In the regularized theory developed here, the torsionful branch occurs when
\begin{equation}
  2\,\alpha_0\,\alpha_2 + 3\,\alpha_1^2 = 0 \,,
\end{equation}
which can be regarded as the regularized counterpart of the above criterion. 
More broadly, it is plausible that, once an $n$-dimensional torsion-branch
criterion is known in a given model in RC geometry, applying the same regularization map to
the couplings yields the corresponding parameter constraint in our framework.

By calculating the Kretschmann scalar built from the Levi--Civita connection,
\begin{equation}
   \mathring{R}_{\alpha\beta\mu\nu}\mathring{R}^{\alpha\beta\mu\nu}
   = \frac{4 \alpha_2^2 c_1^2 + 4 \alpha_1 \alpha_2 c_1 (1 + c_1 r^2)
   + \alpha_1^2 (1 + 2 c_1 r^2 + 6 c_1^2 r^4)}
   {\alpha_2^2 c_1^2 r^4}\,,
\end{equation}
we see that $r=0$ is a curvature singularity except for $c_{1}=-\alpha_{1}/(2\alpha_{2})$, 
in which case
$A^2 = 1/B^2 = 1 - r^2\alpha_{1}/(2\alpha_{2})$, corresponding to a de Sitter solution if $\alpha_{1}/(2\alpha_{2}) > 0$ 
or an anti–de Sitter solution if 
$\alpha_{1}/(2\alpha_{2}) < 0$.
However, one should note that the torsion is non-vanishing, because  now we have
\begin{equation}
   P \;=\; \sqrt{\frac{c_2\,\sqrt{1-\frac{r^2 \alpha_1}{2 \alpha_2}}}{\,r\,}} \,.
\end{equation}
(Anti)–de Sitter solutions with nonzero torsion are not uncommon in torsionful gravity models. 
For example, in the de Sitter gauge theory of gravity, it is shown that among spherically symmetric vacua with line element
\(ds^{2}=A^{2}(r)\,dt^{2}-B^{2}(r)\,dr^{2}-r^{2}(d\theta^{2}+\sin^{2}\theta\,d\phi^{2})\),
the only solution is the static de Sitter spacetime supplemented by a static, \(\mathrm{O}(3)\)-symmetric vector torsion \cite{huang2009sitter}.

For a general \(c_{1}<0\),  an event horizon exists iff \(c_{1}<0\), located at
\[
r_{H}=\frac{1}{\sqrt{-c_{1}}}\, .
\]
The large–\(r\) behavior is then  governed by the \(r^{2}\)-coefficient of \(A^{2}\), namely
\(-\alpha_{1}/(2\alpha_{2})\):
\[
\begin{cases}
\displaystyle \frac{\alpha_{1}}{\alpha_{2}}>0 \; \Rightarrow \; \text{asymptotically de Sitter, and the unique horizon } r_{H} \text{ is cosmological},\\[1.2em]
\displaystyle \frac{\alpha_{1}}{\alpha_{2}}<0 \; \Rightarrow \; \text{asymptotically anti–de Sitter, and } r_{H} \text{ is the black–hole event horizon}.
\end{cases}
\]

Overall, this solution exhibits a long-range torsion field sustained by the GB term. In contrast, in minimal EC theory torsion is nondynamical and vanishes in vacuum, so this solution highlights the role of the GB term in sustaining torsion outside the black hole.

\subsection{Solution 2}

Building on the ansatz structure from Cembranos and Valcarcel \cite{cembranos2017new}, with the same  parameter relation $2\alpha_{0}\alpha_{2}+3\alpha_{1}^2=0$ , we obtain a distinct class of black hole solution characterized by a logarithmic scalar field profile. The solution is given by:
\begin{equation}
\begin{aligned}
 &f_{1}=-\frac{c_1\alpha_1-2c_3\alpha_2+2\alpha_{1}r}{4\alpha_2}\,, \qquad
 p_{1}=\frac{-c_1 r \alpha_1+4 c_1 c_3 \alpha_2+2 c_3 r \alpha_2}{2 \sqrt{2} \alpha_2 \sqrt{-\frac{\left(c_1+r\right)\left(r \alpha_1-2 c_3 \alpha_2\right)}{\alpha_2}}}\,, \\[0.3em]
 & q_{1}=\frac{r\left(c_1 \alpha_1+2 r \alpha_1-2 c_3 \alpha_2\right)}{2 \sqrt{2} \alpha_2 \sqrt{-\frac{\left(c_1+r\right)\left(r \alpha_1-2 c_3 \alpha_2\right)}{\alpha_2}}}\,, \qquad
 f_{2}=g_{1}=g_{2}=p_{2}=q_{2}=0\,.
\end{aligned}
\end{equation}
and 
\begin{equation}
\begin{aligned}
 A^2 &=\frac{1}{B^2}=-\frac{\left(c_1+r\right)\left(r \alpha_1-2 c_3 \alpha_2\right)}{2 \alpha_2}\,, \\[0.3em]
 \phi&=\ln c_2 - \ln\!\bigl|c_1+r\bigr|\,.
\end{aligned}
\end{equation}
The roots for 
$g^{rr}=0$ are $r_{1}=-c_{1}$
and 
$r_{2} = 2c_{3}\alpha_{2}/\alpha_{1}$, meaning that there are at most two event horizons. We can rewrite $A^2=1/B^2$ as 
\begin{equation}
   A^2=\frac{1}{ B^2}=-\frac{\alpha_{1}}{2\alpha_{2}}(r-r_{1})(r-r_{2})\,.
\end{equation}
When $-\alpha_{1}/(2\alpha_{2})>0$, this solution corresponds to an asymptotically anti–de Sitter black hole, with the outer horizon being the black hole event horizon. When $-\alpha_{1}/(2\alpha_{2})<0$, this solution corresponds to an asymptotically de Sitter black hole, with the outer and inner horizons being the cosmological  and the black hole event horizon, respectively. 
One can also check the three independent quadratic torsion invariants and find that they all vanish, i.e.,
\begin{equation}
 T_{\lambda \mu \nu} T^{\lambda \mu \nu}
 \;=\;
 T_{\lambda \mu \nu} T^{\mu \lambda \nu}
 \;=\;
 T^{\mu}{}_{\mu}{}_{\lambda} T^{\nu}{}_{\nu}{}^{\lambda}
 \;=\; 0\,.
\end{equation}

\section{Conclusion}

In this work we have developed a conformal regularization of the Gauss--Bonnet term in four-dimensional Riemann--Cartan geometry based on a consistent dimensional derivative prescription. This scheme circumvents the ambiguities that arise when discarding would-be boundary terms in the presence of torsion, yielding a well-defined four-dimensional action. Unlike the usual ``$\mathcal{G}/(n-4)$ followed by $n\!\to\!4$'' procedure by \cite{glavan2020einstein}, our construction avoids steps that become ill-defined in RC geometry and, in the torsionless sector, reduces smoothly (up to boundary terms) to the scalar--tensor realization of four-dimensional Einstein--Gauss--Bonnet gravity obtained via conformal regularization \cite{fernandes20224d,fernandes2020derivation}. In this sense, our model provides a genuine Riemann--Cartan extension of the regularized 4D EGB theory.

Starting from this action, we have derived the complete field equations by independently varying with respect to the scalar, tetrad, and spin connection. A central structural result is that, although the regularized Gauss--Bonnet density contains second derivatives of the scalar and explicit torsion, all Euler--Lagrange equations of the regularized ECGB theory are strictly second order in covariant derivatives of the dynamical fields. In particular, the three apparently third-derivative terms in the scalar equation can be recombined via the  commutator into $R\nabla\phi$ and $T\nabla\nabla\phi$ structures, so that no derivatives higher than second order act on $\phi$.
Together with the manifestly second-order tetrad and spin-connection equations, this ensures that the theory is free from Ostrogradsky-type instabilities associated with higher-than-second time derivatives. We have also exhibited static, spherically symmetric black hole solutions with nonvanishing torsion. A central outcome is that, even in four dimensions—where the Gauss–Bonnet density is topological—the regularized Gauss–Bonnet coupling acts as an intrinsic source of long-range torsion, allowing black-hole solutions to carry torsion hair. This furnishes a minimal four-dimensional route to long-range torsion, without recourse to extra dimensions.

From the resulting equations of motion, written both in tetrad--connection and coordinate form, we have uncovered new curvature--torsion--scalar couplings with no counterpart in the torsion-free case. Notably, the trace relation differs from the familiar $4\alpha_0+2\alpha_1 R-\alpha_2\mathcal{G}=0$ and reduces to it only when torsion vanishes, highlighting the genuinely RC character of the dynamics. We have also validated our variational procedure by reproducing the two-dimensional Einstein and Einstein–Cartan limits, which provide nontrivial consistency checks of the regularization and of the conformal transformation rules adopted.

Within the static, spherically symmetric sector, we have constructed two representative families of solutions.
Both families lie on the torsionful branch singled out by the regularized coupling constraint $2\,\alpha_{0}\alpha_{2}+3\,\alpha_{1}^{2}=0$, mirroring the situation in higher-dimensional Lovelock--Cartan gravity, where static black holes with nonvanishing torsion arise only at a special coupling tuning (cf. Eq.~(3.8) in Ref.~\cite{cvetkovic2018black}), although the explicit relation differs because our theory is conformally regularized in four dimensions.
Solution~1 carries a nontrivial, radially decaying torsion profile; a special subcase with $c_{1}=-\alpha_{1}/(2\alpha_{2})$ yields a metric of (anti)–de Sitter type while retaining nonzero torsion, in line with known (A)dS solutions with spherically symmetric vector torsion in de Sitter gauge gravity \cite{huang2009sitter}.
The solution~2 features a logarithmic scalar profile, at most two horizons, and vanishing quadratic torsion invariants.
Together, these examples illustrate how the conformally regularized GB interaction can support genuine torsion hair in four dimensions, providing a purely four-dimensional counterpart to previously known torsionful black holes in $D>4$.

\begin{acknowledgments}
The authors would like to thank Changjun Gao for helpful feedback on the manuscript.
This work is supported in part by the National Key 
Research and Development Program of China under Grant No.~2020YFC2201501 
and the National Natural Science Foundation of China (NSFC) under Grant No.~12347103.
\end{acknowledgments}

\appendix
\section{Vanishing of the torsion scalar in two dimensions}
\label{Appendix:vanishing_of_the_torsion_scalar_in_two_dimensions}
It is easy to show that in dimension $n=2$ the most general torsion is purely vectorial,
\begin{equation}
    T^\rho{}_{\mu\nu}=\delta^\rho_\nu\,T_\mu-\delta^\rho_\mu\,T_\nu \;.
\end{equation}
Then, the two quadratic contractions entering $T$ are
\begin{align}
T_{\alpha\beta\gamma}T^{\alpha\beta\gamma}
&=(g_{\alpha\beta}T_\gamma-g_{\alpha\gamma}T_\beta)
  (g^{\alpha\beta}T^\gamma-g^{\alpha\gamma}T^\beta)
=2\,T_\mu T^\mu,\\[0.3em]
T_{\alpha\beta\gamma}T^{\gamma\beta\alpha}
&=(g_{\alpha\beta}T_\gamma-g_{\alpha\gamma}T_\beta)
  (g^{\gamma\beta}T^\alpha-g^{\alpha\gamma}T^\beta)
=\,T_\mu T^\mu \;.
\end{align}
As a result, one has
\begin{equation}
\begin{aligned}
T
&\equiv \frac{1}{4} T_{\alpha \beta \gamma} T^{\alpha \beta \gamma}
      +\frac{1}{2} T_{\alpha \beta \gamma} T^{\gamma \beta \alpha}
      -T_\alpha T^\alpha\\
&=\Bigl[\tfrac{1}{4}\cdot 2+\tfrac{1}{2}-1\Bigr]\,T_\mu T^\mu
=0\,.
\end{aligned}
\end{equation}
 we conclude $T\equiv 0$ identically.

\section{Intermediate steps for the conformal regularization in EC gravity}
\label{app:dim-reg-details}

In this Appendix we collect the intermediate steps leading to the compact expressions
\eqref{eq:dRdn-2d-EC}, \eqref{eq:delta-phi-metric-affine-compact},
\eqref{eq:delta-phi-tetrad-compact}, and \eqref{eq:delta-e-tetrad-compact} used in
Section~\ref{sec:EC-regularized}.

Starting from the conformal transformation of the Ricci scalar with torsion, the
dimensional regularization of $\sqrt{-\tilde{g}}\,\tilde{R}$ yields
\begin{equation}
\begin{aligned}
 \frac{d  \bigl( \sqrt{-\tilde{g}}\, \tilde{R}\bigr) }{dn}\bigg|_{n=2}
 =&\;\sqrt{-g}\left\{\phi R-2(1+\phi)\nabla^2\phi
 -2\phi T_{\alpha}\nabla^{\alpha}\phi-(\nabla \phi)^2\right\}\,,\\[0.3em]
 =&\;\sqrt{-g}\left\{\phi\bigl( \mathring{R}+T+2 \mathring{\nabla}_{\alpha}T^{\alpha}\bigr)
 -2(1+\phi)\bigl(\mathring{\nabla}{}^2
\phi-T^{\mu}\mathring{\nabla}_{\mu}\phi\bigr)
-2\phi T_{\alpha}\nabla^{\alpha}\phi-(\nabla \phi)^2\right\}\,,\\[0.3em]
=&\;\sqrt{-g}\left\{\phi\bigl( \mathring{R}+T\bigr)+(\mathring\nabla \phi)^2\right\}
+\text{boundary term}\,.
\end{aligned}
\end{equation}
These identities make explicit the derivation of the compact expression
\eqref{eq:dRdn-2d-EC} used in the main text.

The variation of the conformally rescaled Ricci scalar with respect to the scalar
field $\phi$ in the metric--affine formulation can be written as
\begin{equation}
\begin{aligned}
   \delta_{\phi}  \bigl(\sqrt{-\tilde{g}}\, \tilde{R}\bigr)
   =&\;\sqrt{-\tilde{g}} \Biggl[\tilde{\mathring{G}}_{\mu\nu}
   +\frac{1}{2} \biggl(
   -\tilde{g}_{\mu\nu}\tilde{K}_{\alpha\gamma\beta}\tilde{K}^{\alpha\beta\gamma}
   -\tilde{g}_{\mu\nu}\tilde{K}^{\alpha\beta}{}_{\alpha}\tilde{K}_{\beta}{}^{\gamma}{}_{\gamma}
   -\tilde{K}_{\alpha\beta\nu}\tilde{K}_{\mu}{}^{\alpha\beta}\\[0.2em]
   &\qquad\qquad
   +\tilde{K}_{\alpha}{}^{\beta}{}_{\beta}\tilde{K}_{\mu}{}^{\alpha}{}_{\nu}
   -\tilde{K}_{\alpha\beta\mu}\tilde{K}_{\nu}{}^{\alpha\beta}
   +\tilde{K}_{\alpha}{}^{\beta}{}_{\beta}\tilde{K}_{\nu}{}^{\alpha}{}_{\mu}
   \biggr)\Biggr] \bigl(-2\tilde{g}^{\mu\nu}\bigr)\delta\phi\\
   =&\;(n-2)\bigl(\tilde{T}+\tilde{\mathring{R}}\bigr)\sqrt{-\tilde{g}}\,\delta\phi\,,
\end{aligned}
\end{equation}
which is the full form corresponding to the compact expression
\eqref{eq:delta-phi-metric-affine-compact} quoted in the main text.

In the tetrad--spin-connection formulation, the variation of
$\sqrt{-\tilde{g}}\,\tilde{R}$ with respect to $\phi$ can be organized as
\begin{equation}
\begin{aligned}
       \delta_{\phi} \bigl(\sqrt{-\tilde{g}}\, \tilde{R}\bigr)
       =&\;\sqrt{-g} \Bigl[-2\tilde{R}^{\mu }{}_{a}
       +\tilde{R}\,\tilde{e}_{a}{}^{\mu}\Bigr]\exp(n\phi)\,  \tilde{e}^{a} {}_{\mu}\,\delta \phi\\
       &\;+\sqrt{-g}\,\exp(n\phi)\Bigl[\tilde{T}^{\mu}{}_{ab}
       +\tilde{T}_{b}\,\tilde{e}_{a}{}^{\mu}
       -\tilde{T}_{a}\,\tilde{e}_{b}{}^{\mu}\Bigr]
       \Bigl[e^a{}_{\mu}e^{b\nu}\partial_{\nu}\delta \phi
       -e^b{}_{\mu}e^{a\nu}\partial_{\nu}\delta \phi \Bigr]\\
       =&\;\sqrt{-g}\,(n-2)\tilde{R}\,\exp(n\phi)\,\delta \phi\\
       &\;+\sqrt{-g}\,\exp\!\bigl((n-2)\phi\bigr)\Bigl[{T}^{\mu}{}_{ab}
       +{T}_{b}{e}_{a}{}^{\mu}-{T}_{a}{e}_{b}{}^{\mu}\Bigr]
       \Bigl[e^a{}_{\mu}e^{b\nu}\partial_{\nu}\delta \phi
       -e^b{}_{\mu}e^{a\nu}\partial_{\nu}\delta \phi \Bigr]\\
       =&\;\sqrt{-g}\,(n-2)\tilde{R}\,\exp(n\phi)\,\delta \phi
       +\sqrt{-g}\,\exp\!\bigl((n-2)\phi\bigr)\,(2n-4)\,T^{\nu}\mathring{\nabla}_{\nu}\delta \phi\\
       =&\;\sqrt{-g}\,(n-2)\tilde{R}\,\exp(n\phi)\,\delta \phi
       -\sqrt{-g}\,\mathring{\nabla}_{\nu}\Bigl[\exp\!\bigl((n-2)\phi\bigr)\,(2n-4)\,T^{\nu}\Bigr]\delta \phi\\
       &\;+\text{boundary terms}\,,
\end{aligned}
\end{equation}
which yields the compact result~\eqref{eq:delta-phi-tetrad-compact} used in
Section~\ref{sec:EC-regularized}.

Similarly, the variation with respect to the tetrad field $e^{a}{}_{\mu}$ proceeds as
\begin{equation}
\begin{aligned}
      & \delta_{e} \bigl(\sqrt{-\tilde{g}}\, \tilde{R}\bigr)\\
       =&\;\sqrt{-g} \Bigl[-2\tilde{R}^{\mu }{}_{a}
       +\tilde{R}\,\tilde{e}_{a}{}^{\mu}\Bigr]\exp\!\bigl((n+1)\phi\bigr)\delta e^{a}{}_{\mu}\\
       &\;+\sqrt{-g}\,\exp(n\phi)\Bigl[\tilde{T}^{\mu}{}_{ab}
       +\tilde{T}_{b}\,\tilde{e}_{a}{}^{\mu}
       -\tilde{T}_{a}\,\tilde{e}_{b}{}^{\mu}\Bigr]
       \Bigl[e^{b\nu}\partial_{\nu}\phi\,  \delta  e^{a}{}_{\mu}
       - e^{a}{}_{\mu}e^{\nu}{}_{d}e^{b\alpha}\partial_{\nu}\phi\,\delta e^{d}{}_{\alpha}
       -(a\leftrightarrow b)\Bigr]\\
       =&\;\sqrt{-g} \Bigl[-2\tilde{R}^{\mu }{}_{\nu}
       +\tilde{R}\,\delta_{\nu}^{\mu}\Bigr]e_{a}{}^{\nu}\exp(n\phi)\,\delta e^{a}{}_{\mu}\\
       &\;+\sqrt{-g}\,\exp\!\bigl((n-2)\phi\bigr)\Bigl[
       2T^{\mu}{}_{a\beta} \mathring{\nabla}{}^{\beta}\phi
       +2e_{a}{}^{\mu}T_\beta \mathring{\nabla}{}^{\beta}\phi
       -2T_{a}\mathring{\nabla}{}^{\mu}\phi
       -2(n-2)T^{\mu}\mathring{\nabla}{}_{a}\phi\Bigr]\delta e^{a}{}_{\mu}\,.
\end{aligned} 
\end{equation}
This chain of equalities provides the detailed expression corresponding to the
compact formula \eqref{eq:delta-e-tetrad-compact} employed in
Section~\ref{sec:EC-regularized}.

\section{Equations of motion}
\label{Appendix:equations_of_motion}
For completeness, we collect the full equations of motion obtained from the conformally regularized action in four-dimensional RC geometry. In the torsion-free limit ($K^\lambda{}_{\mu\nu}=0$, $T^\lambda{}_{\mu\nu}=0$), these expressions are consistently reduced to the familiar four-dimensional {EGB} equations obtained via the conformal/dimensional regularization; see, e.g., Ref.~\cite{fernandes20224d}.

\paragraph{Scalar equation.}
Varying the action with respect to the scalar $\phi$ yields  Eqs.~\eqref{4d_ec_phi}:
\begin{align}
\label{tensor_equation_final}
& \mathcal{G} +8T^{\gamma}T_{\gamma}\nabla^{2}\phi-4R\nabla^{2}\phi-4T^{\beta\gamma\kappa}R_{\gamma\kappa\alpha\beta}\nabla^{\alpha}\phi+8T^{\gamma}R_{\gamma\alpha}\nabla^{\alpha}\phi +8T^{\beta}{}_{\alpha}{}^{\gamma}R_{\gamma}{}_{\beta}\nabla^{\alpha}\phi\nonumber\\
&-4T_{\alpha}R\nabla^{\alpha}\phi-4R\nabla_{\beta}T^{\beta}+8R_{\gamma\alpha}\nabla_{\beta}T^{\alpha\beta\gamma}+8(\nabla^{2}\phi)^2+16T_{\alpha}\nabla^{\alpha}\phi\nabla^{2}\phi+8(\nabla\phi)^2\nabla^2\phi\nonumber\\
&-8T_{\alpha}\nabla_{\beta}\nabla^{\beta}\nabla^{\alpha}\phi+8T_{\alpha}T_{\beta}\nabla^{\alpha}\phi\nabla^{\beta}\phi-8T_{\alpha\beta}{}^{\gamma}T_{\gamma}\nabla^{\alpha}\phi\nabla^{\beta}\phi-8R_{\alpha\beta}\nabla^{\alpha}\phi\nabla^{\beta}\phi+8T_{\beta}(\nabla\phi)^2\nabla^{\beta}\phi\nonumber\\&+8\nabla^{\alpha}\phi\nabla_{\beta}T_{\alpha}\nabla^{\beta}\phi+16\nabla^{\alpha}\phi\nabla_{\beta}\nabla_{\alpha}\phi\nabla^{\beta}\phi+8T_{\beta}\nabla^{\beta}\nabla^2\phi-8T_{\alpha}T_{\beta}\nabla^{\beta}\nabla^{\alpha}\phi\nonumber\\&+8T_{\beta\alpha}{}^{\gamma}T_{\gamma}\nabla^{\beta}\nabla^{\alpha}\phi+8R_{\beta\alpha}\nabla^{\beta}\nabla^{\alpha}\phi-8\nabla_{\alpha}\nabla_{\beta}\phi\nabla^{\beta}\nabla^{\alpha}\phi-8\nabla_{\beta}T_{\alpha}\nabla^{\beta}\nabla^{\alpha}\phi\nonumber\\&-8\nabla^{\alpha}\phi\nabla^{\beta}\phi\nabla_{\gamma}T_{\alpha\beta}{}^{\gamma}+8\nabla^{\beta}\nabla^{\alpha}\phi \nabla_{\gamma}T_{\beta\alpha}{}^{\gamma}+8\nabla_{\alpha}\nabla^{\alpha}\phi \nabla_{\gamma}T^{\gamma}-8T^{\alpha\beta\gamma}\nabla_{\gamma}R_{\beta\alpha}\nonumber\\&+8R_{\beta\gamma}\nabla^{\gamma}T^{\beta}-8T_{\alpha\beta\gamma}\nabla^{\alpha}\phi\nabla^{\gamma}\nabla^{\beta}\phi-8T_{\beta\alpha\gamma}\nabla^{\alpha}\phi\nabla^{\gamma}\nabla^{\beta}\phi-8T_{\gamma\alpha\beta}\nabla^{\alpha}\phi\nabla^{\gamma}\nabla^{\beta}\phi\nonumber\\&+8T_{\beta\alpha\gamma}\nabla^{\gamma}\nabla^{\beta}\nabla^{\alpha}\phi+4T^{\alpha\beta\gamma}\nabla_{\kappa}R_{\beta\gamma\alpha}{}^{\kappa}-4T^{\beta}\Big(-2T^{\kappa}R_{\beta\kappa}+T^{\gamma\kappa\mu}R_{\kappa\mu\beta\gamma}-2T^{\gamma}{}_{\beta}{}^{\kappa}R_{\kappa\gamma}\nonumber\\&+T_{\beta}R+\nabla_{\beta}R-2\nabla_{\kappa}R_{\beta}{}^{\kappa}\Big)+4R_{\beta\gamma\alpha\kappa}\nabla^{\kappa}T^{\alpha\beta\gamma}=0\;.
\end{align}
This formula contains three terms that involve third-order covariant derivatives of the scalar field $\phi$. 
In this appendix we show explicitly that their sum can be rewritten in terms of at most second-order covariant derivatives of $\phi$, 
so that no Ostrogradsky instability arises from these contributions.  
The potentially dangerous terms are
\begin{equation}
 -8T_{\alpha}\nabla_{\beta}\nabla^{\beta}\nabla^{\alpha}\phi  \;,
 \quad
 8T_{\beta}\nabla^{\beta}\nabla^2\phi\;, 
 \quad\text{and}\quad
 8T_{\beta\alpha\gamma}\nabla^{\gamma}\nabla^{\beta}\nabla^{\alpha}\phi\;.
\end{equation}

We first consider the sum of the first two terms, which is
\begin{equation}
\begin{aligned}
&\quad -8T_{\alpha}\nabla_{\beta}\nabla^{\beta}\nabla^{\alpha}\phi
          + 8T_{\beta}\nabla^{\beta}\nabla^2\phi \\
&     = 8 T_\beta\left[\nabla^\beta, \nabla_\alpha \nabla^\alpha\right] \phi \\
&     = 8T_\beta \left(\left[\nabla^\beta, \nabla_\alpha\right] \nabla^\alpha \phi
          +\nabla_\alpha\left[\nabla^\beta, \nabla^\alpha\right]\phi\right) \\
&     = 8T_\beta\Bigl(
        -R_{\sigma}{}^{\beta}\,\nabla^{\sigma}\phi
        - T^{\lambda\beta}{}_{\alpha}\,\nabla_{\lambda}\nabla^{\alpha}\phi
        - \nabla_{\alpha}T^{\lambda\beta\alpha}\,\nabla_{\lambda}\phi
        - T^{\lambda\beta\alpha}\,\nabla_{\alpha}\nabla_{\lambda}\phi
        \Bigr)\,.
\end{aligned}
\end{equation}

Here we have used the commutator relations in Riemann--Cartan geometry,
\begin{equation}
    \left[\nabla_\mu, \nabla_\nu\right] \phi=-T^\lambda{ }_{\mu \nu} \nabla_\lambda \phi
\end{equation}
and
\begin{equation}
    \left[\nabla_\mu, \nabla_\nu\right] V^\rho=R^\rho{ }_{\sigma \mu \nu} V^\sigma-T^\lambda{ }_{\mu \nu} \nabla_\lambda V^\rho
\end{equation}
for an arbitrary vector $V^{\mu}$.
For the third term we proceed in an analogous way {resulting in}
\begin{equation}
\begin{aligned}
8T_{\beta\alpha\gamma}\nabla^{\gamma}\nabla^{\beta}\nabla^{\alpha}\phi
&= 4T_{\beta\alpha\gamma}
   \bigl(\nabla^{\gamma}\nabla^{\beta}\nabla^{\alpha}\phi
        -\nabla^{\alpha}\nabla^{\beta}\nabla^{\gamma}\phi\bigr)
\\
&= 4T_{\beta\alpha\gamma}
   \bigl([\nabla^{\beta},\nabla^{\alpha}]\nabla^{\gamma}\phi
        +\nabla^{\beta}[\nabla^{\gamma},\nabla^{\alpha}]\phi
        +[\nabla^{\gamma},\nabla^{\beta}]\nabla^{\alpha}\phi\bigr)
\\
&= 4T_{\beta\alpha\gamma}\Bigl(
    R^{\gamma}{}_{\sigma}{}^{\beta\alpha}\nabla^{\sigma}\phi
   -T^{\rho\beta\alpha}\nabla_{\rho}\nabla^{\gamma}\phi
   -(\nabla^{\beta}T^{\rho\gamma\alpha})\nabla_{\rho}\phi
\\
&\qquad\qquad\quad
   {-T^{\rho\gamma\alpha}\nabla^{\beta}\nabla_{\rho}\phi}
   +R^{\alpha}{}_{\sigma}{}^{\gamma\beta}\nabla^{\sigma}\phi
   -T^{\rho\gamma\beta}\nabla_{\rho}\nabla^{\alpha}\phi
   \Bigr)\, .
\end{aligned}
\end{equation}
Combining these results, we see that the whole combination of terms
with three covariant derivatives of $\phi$ can be rewritten entirely in
terms of curvature and torsion tensors multiplying at most second-order
covariant derivatives of $\phi$ (plus first derivatives and undifferentiated
fields). 
Therefore, although the intermediate expressions contain $\nabla\nabla\nabla\phi$,
{we have proved that}
the final scalar field equation remains strictly second order in derivatives.

\paragraph{Tetrad  equation.}
Variation with respect to the tetrad  $\omega^{ab}{}_{\mu}$ leads to Eq \eqref{4d_ec_metric} :
\begin{align}
\label{new_tensor_equation_balanced}
& -2 \alpha_1 R_{\mu}{}^{\nu} + \delta_{\mu}{}^{\nu} \Big( \alpha_0 + \alpha_1 R + R (-4 \alpha_2 \nabla^{\alpha}\nabla_{\alpha}\phi + 2 \alpha_2 \nabla^{\alpha}\phi \nabla_{\alpha}\phi) \nonumber\\
& \qquad + 4 \alpha_2 (\nabla^{\alpha}\nabla_{\alpha}\phi) (\nabla^{\beta}\nabla_{\beta}\phi) - 8 \alpha_2 R^{\alpha\beta} \nabla_{\alpha}\phi \nabla_{\beta}\phi - 2 \alpha_2 (\nabla^{\alpha}\phi \nabla_{\alpha}\phi) (\nabla^{\beta}\phi \nabla_{\beta}\phi) \nonumber\\
& \qquad + 8 \alpha_2 \nabla_{\alpha}\phi \nabla^{\beta}(\nabla^{\alpha}\phi) \nabla_{\beta}\phi + 8 \alpha_2 R^{\beta\alpha} \nabla_{\beta}\nabla_{\alpha}\phi - 4 \alpha_2 \nabla^{\alpha}\nabla^{\beta}\phi \nabla_{\beta}\nabla_{\alpha}\phi \Big) \nonumber\\
& - 4 \alpha_2 \Big( R_{\mu}{}^{\nu} (-2 \nabla^{\alpha}\nabla_{\alpha}\phi + \nabla^{\alpha}\phi \nabla_{\alpha}\phi) - 2 R_{\mu}{}^{\alpha\nu\beta} \nabla_{\alpha}\phi \nabla_{\beta}\phi + 2 R_{\mu}{}^{\beta\nu\alpha} \nabla_{\beta}(\nabla_{\alpha}\phi) \nonumber\\
& \qquad + 2 R^{\alpha\nu} (\nabla_{\alpha}\nabla_{\mu}\phi - \nabla_{\alpha}\phi \nabla_{\mu}\phi) + T_{\alpha\beta\gamma} R^{\beta\gamma}{}_{\mu}{}^{\alpha} \nabla^{\nu}\phi - 2 T_{\alpha\mu\beta} R^{\beta\alpha} \nabla^{\nu}\phi \nonumber\\
& \qquad - 2 T_{\beta} R^{\beta}{}_{\mu} \nabla^{\nu}\phi + T_{\mu} R\nabla^{\nu}\phi - 2 T_{\mu}(\nabla^{\alpha}\nabla_{\alpha}\phi) \nabla^{\nu}\phi \nonumber\\
& \qquad - 2 R_{\mu}{}^{\alpha} \nabla_{\alpha}\phi \nabla^{\nu}\phi + 2 (\nabla^{\alpha}\nabla_{\mu}\phi) \nabla_{\alpha}\phi \nabla^{\nu}\phi - 2 T^{\alpha}{}_{\mu}{}^{\beta} \nabla_{\alpha}\phi \nabla_{\beta}\phi \nabla^{\nu}\phi \nonumber\\
& \qquad + 2 T^{\beta}{}_{\mu}{}^{\alpha} \nabla_{\beta}(\nabla_{\alpha}\phi) \nabla^{\nu}\phi + R \nabla_{\mu}\phi \nabla^{\nu}\phi - 2 (\nabla^{\alpha}\nabla_{\alpha}\phi) \nabla_{\mu}\phi \nabla^{\nu}\phi \nonumber\\
& \qquad - 2 T^{\alpha} \nabla_{\alpha}\phi \nabla_{\mu}\phi \nabla^{\nu}\phi - 2 (\nabla^{\alpha}\phi \nabla_{\alpha}\phi) \nabla_{\mu}\phi \nabla^{\nu}\phi + 2 T^{\alpha} \nabla_{\mu}\nabla_{\alpha}\phi \nabla^{\nu}\phi \nonumber\\
& \qquad - 2 \nabla_{\alpha}\nabla_{\mu}\phi \nabla^{\nu}\nabla^{\alpha}\phi + 2 \nabla_{\alpha}\phi \nabla_{\mu}\phi\nabla^{\nu}\nabla^{\alpha}\phi + 2 R_{\mu}{}^{\alpha} \nabla^{\nu}\nabla_{\alpha}\phi\nonumber\\
& \qquad - R\nabla^{\nu}\nabla_{\mu}\phi + 2 \nabla^{\alpha}\nabla_{\alpha}\phi \nabla^{\nu}\nabla_{\mu}\phi \Big)=0\,.
\end{align}

\paragraph{Spin connection equation.}
Variation with respect to the independent spin connection $\omega^{ab}{}_{\mu}$ leads to Eq \eqref{4d_ec_spin} :
\begin{align}
\label{torsion_equation_2}
& \alpha_1 T_{\mu}{}^{\alpha\beta} - 4 \alpha_2 T_{\mu}{}^{\beta\gamma} \nabla^{\alpha}\nabla_{\gamma}\phi + 4 \alpha_2 T^{\beta} \nabla^{\alpha}\nabla_{\mu}\phi + 4 \alpha_2 T_{\mu}{}^{\alpha\gamma} \nabla^{\beta}\nabla_{\gamma}\phi \nonumber\\
& - 4 \alpha_2 T_{\gamma}{}^{\alpha\gamma} \nabla^{\beta}\nabla_{\mu}\phi - 4 \alpha_2 T_{\mu}{}^{\alpha\beta} \nabla^{\gamma}\nabla_{\gamma}\phi + 4 \alpha_2 T_{\mu}{}^{\beta\gamma} \nabla^{\alpha}\phi \nabla_{\gamma}\phi \nonumber\\
& - 4 \alpha_2 T_{\mu}{}^{\alpha\gamma} \nabla^{\beta}\phi \nabla_{\gamma}\phi + 2 \alpha_2 T_{\mu}{}^{\alpha\beta} \nabla^{\gamma}\phi \nabla_{\gamma}\phi + 4 \alpha_2 T^{\gamma\alpha\beta} \nabla_{\gamma}\nabla_{\mu}\phi \nonumber\\
& - \delta^{\beta}{}_{\mu} \Big( \alpha_1 T^{\alpha} + 2 \alpha_2 \big( 2 T^{\gamma} (\nabla^{\alpha}\nabla_{\gamma}\phi - \nabla^{\alpha}\phi \nabla_{\gamma}\phi) \nonumber\\
& \qquad\qquad + T^{\alpha} (-2 \nabla^{\gamma}\nabla_{\gamma}\phi + \nabla^{\gamma}\phi \nabla_{\gamma}\phi) - 2 T^{\gamma\alpha\kappa} \nabla_{\gamma}\phi \nabla_{\kappa}\phi + 2 T^{\kappa\alpha\gamma} \nabla_{\kappa}\nabla_{\gamma}\phi \big) \Big) \nonumber\\
& + \delta^{\alpha}{}_{\mu} \Big( \alpha_1 T^{\beta} + 2 \alpha_2 \big( 2 T^{\gamma} (\nabla^{\beta}\nabla_{\gamma}\phi - \nabla^{\beta}\phi \nabla_{\gamma}\phi) \nonumber\\
& \qquad\qquad + T_{\kappa}{}^{\beta\kappa} (-2 \nabla^{\gamma}\nabla_{\gamma}\phi + \nabla^{\gamma}\phi \nabla_{\gamma}\phi) - 2 T^{\gamma\beta\kappa} \nabla_{\gamma}\phi \nabla_{\kappa}\phi + 2 T^{\kappa\beta\gamma} \nabla_{\kappa}\nabla_{\gamma}\phi \big) \Big) \nonumber\\
& - 4 \alpha_2 T^{\beta} \nabla^{\alpha}\phi \nabla_{\mu}\phi + 4 \alpha_2 T^{\alpha} \nabla^{\beta}\phi \nabla_{\mu}\phi - 4 \alpha_2 T^{\gamma\alpha\beta} \nabla_{\gamma}\phi \nabla_{\mu}\phi=0\,.
\end{align}

\paragraph{Trace equation.}
Contracting the tetrad equation and combining, when useful, with the scalar equation yields the following trace identity, which is reduced to the well-known torsion-free relation when $T^\lambda{}_{\mu\nu}=0$:
\begin{align}
\label{regular-ecg-trace-eq}
& 4 \alpha_0 - \alpha_2\mathcal{G} + 2 \alpha_1 R \nonumber\\
& - 8 \alpha_2 T_{\gamma} T^{\gamma} \nabla^{\alpha}\nabla_{\alpha}\phi - 8 \alpha_2 T^{\beta} \nabla^{\alpha}(\nabla_{\beta}\phi) \nabla_{\alpha}\phi + 4 \alpha_2 R \nabla^{\beta}T_{\beta} \nonumber\\
& - 8 \alpha_2 R^{\gamma\alpha} \nabla^{\beta}(T_{\alpha\beta\gamma}) - 8 \alpha_2 T^{\alpha} \nabla_{\alpha}\phi \nabla^{\beta}\nabla_{\beta}\phi + 8 \alpha_2 T^{\alpha} \nabla^{\beta}(\nabla_{\beta}\nabla_{\alpha}\phi) \nonumber\\
& - 8 \alpha_2 T^{\alpha} T^{\beta} \nabla_{\alpha}\phi \nabla_{\beta}\phi + 8 \alpha_2 T^{\alpha\beta}{}_{\gamma} T^{\gamma} \nabla_{\alpha}\phi \nabla_{\beta}\phi - 8 \alpha_2 \nabla_{\alpha}\phi \nabla^{\beta}T^{\alpha} \nabla_{\beta}\phi \nonumber\\
& - 8 \alpha_2 T^{\beta} \nabla_{\beta}(\nabla^{\alpha}\nabla_{\alpha}\phi) + 8 \alpha_2 T^{\alpha} T^{\beta} \nabla_{\beta}\nabla_{\alpha}\phi - 8 \alpha_2 T^{\beta\alpha}{}_{\gamma} T^{\gamma} \nabla_{\beta}\nabla_{\alpha}\phi \nonumber\\
& + 8 \alpha_2 \nabla^{\beta}(T^{\alpha}) \nabla_{\beta}(\nabla_{\alpha}\phi) + 8 \alpha_2 \nabla_{\alpha}\phi \nabla_{\beta}\phi \nabla^{\gamma}(T^{\alpha\beta}{}_{\gamma}) - 8 \alpha_2 \nabla_{\beta}(\nabla_{\alpha}\phi) \nabla^{\gamma}(T^{\beta\alpha}{}_{\gamma})\nonumber \\
& - 8 \alpha_2 \nabla^{\alpha}\nabla_{\alpha}\phi \nabla^{\gamma}T_{\gamma} + 8 \alpha_2 T_{\alpha\beta\gamma} \nabla^{\gamma}R^{\beta\alpha} - 8 \alpha_2 R^{\beta\gamma} \nabla_{\gamma}T_{\beta} \nonumber\\
& + 8 \alpha_2 T^{\alpha\beta\gamma} \nabla_{\alpha}\phi \nabla_{\gamma}(\nabla_{\beta}\phi) + 8 \alpha_2 T^{\beta\alpha\gamma} \nabla_{\alpha}\phi \nabla_{\gamma}(\nabla_{\beta}\phi) - 8 \alpha_2 T^{\beta\alpha\gamma} \nabla_{\gamma}(\nabla_{\beta}\nabla_{\alpha}\phi)\nonumber \\
& - 4 \alpha_2 T_{\alpha\beta\gamma} \nabla^{\kappa}(R^{\beta\gamma\alpha}{}_{\kappa}) - 4 \alpha_2 R^{\beta\gamma\alpha\kappa} \nabla_{\kappa}(T_{\alpha\beta\gamma}) \nonumber\\
& + 4 \alpha_2 T_{\beta} \Big( -2 T_{\kappa} R^{\beta\kappa} + T_{\gamma\kappa\mu} R^{\kappa\mu\beta\gamma} - 2 T_{\gamma}{}^{\beta}{}_{\kappa} R^{\kappa\gamma} \nonumber\\
& \qquad\qquad\quad + T^{\beta} R + \nabla^{\beta}R - 2 \nabla^{\kappa}R^{\beta}{}_{\kappa} \Big)=0\,.
\end{align}

\section{Reduced Lagrangian for the spherical ansatz}
\label{Appendix:reduced_lagrangian_for_the_spherical_ansatz}
Throughout this section, a prime denotes derivative with respect to $r$, i.e.~$f'(r)\equiv df/dr$.
The reduced Lagrangian entering $\mathcal S=4\pi\!\int dt\,dr\,L$ reads
\begin{align}
\label{eq:Lreduced}
L=\frac{1}{B(r)^4} \bigg\{ & A(r) \bigg[ B(r)^5 \left(r^2 \alpha_0 + 2 \alpha_1 + 2 \alpha_1 p_1(r)^2 - 2 \alpha_1 q_1(r)^2\right) \nonumber\\
& \qquad + 4 r \alpha_1 B(r)^4 \left(g_1(r) p_1(r) + q_1'(r)\right) - 4 r^2 \alpha_2 (11 + 6 \phi(r)) B'(r) \phi'(r)^3 \nonumber\\
& \qquad + 2 r \alpha_2 B(r) \phi'(r)^2 \Big( 8 q_1(r) (3 + 2 \phi(r)) B'(r) + 4 (3 + 2 \phi(r)) \phi'(r) \nonumber\\
& \qquad \qquad + 3 r \phi'(r)^2 + 2 r (11 + 6 \phi(r)) \phi''(r) \Big) \nonumber\\
& \qquad + 4 \alpha_2 B(r)^3 \Big( 4 g_1(r) p_1(r) q_1(r) \phi'(r) - 4 p_1(r) \phi(r) p_1'(r) \phi'(r) \nonumber\\
& \qquad \qquad + 4 q_1(r) (1+\phi(r)) q_1'(r) \phi'(r) - \phi'(r)^2 + 3 q_1(r)^2 \phi'(r)^2 \nonumber\\
& \qquad \qquad - 2(1+\phi(r)) \phi''(r) + 2 q_1(r)^2 (1+\phi(r)) \phi''(r) \nonumber\\
& \qquad \qquad - p_1(r)^2 (\phi'(r)^2 + 2 (1 + \phi(r)) \phi''(r)) \Big) \nonumber\\
& \qquad - 8 \alpha_2 B(r)^2 \phi'(r) \Big( -(1 + p_1(r)^2 - q_1(r)^2) (1 + \phi(r)) B'(r) \nonumber\\
& \qquad \qquad + r (g_1(r) p_1(r) + (3 + 2 \phi(r)) q_1'(r)) \phi'(r) \nonumber\\
& \qquad \qquad + 2 q_1(r) ( (1 + \phi(r)) \phi'(r) + r \phi'(r)^2 + r (3 + 2 \phi(r)) \phi''(r)) \Big) \bigg] \nonumber\\
& + 2 B(r) \bigg[ 2 r \alpha_1 B(r)^4 f_1(r) q_1(r) \nonumber\\
& \qquad\qquad + 2 r^2 \alpha_2 (3 + 2 \phi(r)) \phi'(r)^2 (-2 f_1(r) B'(r) + A'(r) \phi'(r)) \nonumber\\
& \qquad\qquad + B(r)^3 \Big( -(r^2 \alpha_1 + 4 \alpha_2 (1 + p_1(r)^2 - q_1(r)^2) \phi(r)) f_1'(r) \nonumber\\
& \qquad\qquad \qquad + 4 \alpha_2 f_1(r) (-2 p_1(r) \phi(r) p_1'(r) + 2 q_1(r) \phi(r) q_1'(r) \nonumber\\
& \qquad\qquad \qquad \qquad - \phi'(r) - p_1(r)^2 \phi'(r) + 3 q_1(r)^2 \phi'(r)) \Big) \nonumber\\
& \qquad\qquad + 2 r \alpha_2 B(r) \phi'(r) \Big( (-4 q_1(r) (1 + \phi(r)) A'(r) + r (3 + 2 \phi(r)) f_1'(r)) \phi'(r) \nonumber\\
& \qquad\qquad \qquad + 2 f_1(r) (2 q_1(r) (1 + \phi(r)) B'(r) + 2 (1 + \phi(r)) \phi'(r) \nonumber\\
& \qquad\qquad \qquad \qquad + r \phi'(r)^2 + r (3 + 2 \phi(r)) \phi''(r)) \Big) \nonumber\\
& \qquad\qquad - 4 \alpha_2 B(r)^2 \bigg( (2 r q_1(r) f_1'(r) + \phi(r) ((1 + p_1(r)^2 - q_1(r)^2) A'(r) \nonumber\\
& \qquad\qquad\qquad\qquad \qquad + 2 r q_1(r) f_1'(r))) \phi'(r) + f_1(r) \big(2 r (1 + \phi(r)\big) q_1'(r) \phi'(r) \nonumber\\
& \qquad\qquad \qquad\qquad\qquad + q_1(r) \big(r (3 \phi'(r)^2 + 2 \phi''(r)\big)  + 2 \phi(r) (\phi'(r) + r \phi''(r)))) \bigg) \bigg] \bigg\}\,.
\end{align}

For the static and spherically symmetric ansatz introduced in the main text, the above field equations are reduced to a system of ordinary differential equations for the functions $A(r)$, $B(r)$, and $\phi(r)$, together with the nonvanishing spin connection components $f_1(r)$, $g_1(r)$, $p_1(r)$, and $q_1(r)$. 

\paragraph{Equation for $A(r)$.}
\begin{align}
 & -\frac{8 \alpha_2 \phi'(r) B'(r) \left(-p_1(r)^2+q_1(r)^2-1\right)}{B(r)^2} \nonumber\\[0.3em]
& -\frac{8 \alpha_2 r \phi'(r) \left(g_1(r) p_1(r) \phi'(r)+q_1'(r) \phi'(r)+2 q_1(r) \phi''(r)\right)}{B(r)^2} \nonumber\\[0.3em]
& + \frac{2 \alpha_2 r \phi'(r)^2 \left(8 q_1(r) B'(r)+4 r \phi''(r)-r \phi'(r)^2\right)}{B(r)^3} - \frac{8 \alpha_2 r^2 B'(r) \phi'(r)^3}{B(r)^4} \nonumber\\[0.3em]
& + \frac{4 \alpha_2 \left(4 g_1(r) p_1(r) q_1(r) \phi'(r)+\left(p_1(r)^2+1\right) \left(\phi'(r)^2-2 \phi''(r)\right)\right)}{B(r)} \nonumber\\[0.3em]
& + \frac{4 \alpha_2 \left(4 q_1(r) q_1'(r) \phi'(r)+q_1(r)^2 \left(2 \phi''(r)+\phi'(r)^2\right)\right)}{B(r)} \nonumber\\[0.3em]
& + B(r) \left(2 \alpha_1+2 \alpha_1 p_1(r)^2-2 \alpha_1 q_1(r)^2+\alpha_0 r^2\right) \nonumber\\[0.3em]
& + 4 \alpha_1 r \left(g_1(r) p_1(r)+q_1'(r)\right)=0  \,.     
\end{align}

\paragraph{Equation for $B(r)$.}
\begin{align}
 & \frac{1}{B(r)^4} \bigg\{ 8 \alpha_2 B(r)^2 \phi'(r) \left(A'(r) \left(-p_1(r)^2+q_1(r)^2-1\right)+f_1(r) q_1(r) \left(r \phi'(r)-2\right)\right) \nonumber\\
& \qquad\qquad+ 16 \alpha_2 r B(r) \phi'(r)^2 \left(f_1(r)-q_1(r) A'(r)\right) + 8 \alpha_2 r^2 A'(r) \phi'(r)^3 + 4 \alpha_1 r B(r)^4 f_1(r) q_1(r) \nonumber\\
& \qquad\qquad+ A(r) \bigg[ -4 \alpha_2 B(r)^2 \phi'(r) \Big(4 g_1(r) p_1(r) q_1(r) +4 p_1(r) p_1'(r) \nonumber\\
& \qquad\qquad\qquad\qquad\qquad\qquad\qquad\qquad +p_1(r)^2 \phi'(r) +\left(q_1(r)^2+1\right) \phi'(r)\Big) \nonumber\\
& \qquad\qquad\qquad\qquad + 16 \alpha_2 B(r) \phi'(r)^2 (r g_1(r) p_1(r)-q_1(r)) \nonumber\\
& \qquad\qquad\qquad\qquad+ B(r)^4 \left(2 \alpha_1+2 \alpha_1 p_1(r)^2-2 \alpha_1 q_1(r)^2+\alpha_0 r^2\right) \nonumber\\
& \qquad\qquad\qquad\qquad + 2 \alpha_2 r \phi'(r)^3 \left(3 r \phi'(r)+8\right) \bigg] \bigg\}=0 \,.    
\end{align}

\paragraph{Equation for $\phi(r)$.}
\begin{align}
-\frac{1}{B(r)^4} \bigg\{ & 8 \alpha_2 B(r)^4 q_1(r) \left(q_1(r) f_1'(r)+2 f_1(r) q_1'(r)\right) \nonumber\\
& +3 r B'(r) \phi'(r)^2 \left(r A'(r)+A(r) \left(2+r \phi'(r)\right)\right) \nonumber\\
& +B(r)^3 \bigg[ 4 p_1(r) A'(r) p_1'(r)+2 A(r) p_1'(r)^2+2 f_1(r) q_1'(r) \nonumber\\
& \qquad\qquad +2 g_1(r) \left(A(r) q_1(r) p_1'(r)+p_1(r) \left(q_1(r) A'(r)+A(r) q_1'(r)\right)\right) \nonumber\\
& \qquad\qquad +\phi'(r) \left(A'(r)+p_1(r)^2 A'(r)+2 A(r) p_1(r) p_1'(r)-2 r f_1(r) q_1'(r)\right) \nonumber\\
& \qquad\qquad +A''(r) \left(1+p_1(r)^2\right)+2 A(r) p_1(r) p_1''(r)+A(r) \phi''(r) \left(1+p_1(r)^2\right) \nonumber\\
& \qquad\qquad +q_1(r)^2 \left(A'(r) \phi'(r)-A''(r)+A(r) \phi''(r)\right) \nonumber\\
& \qquad\qquad -2 q_1(r) \Big( A'(r) q_1'(r)+f_1(r) \phi'(r)+f_1'(r) \left(-1+r \phi'(r)\right) \nonumber\\
& \qquad \qquad -A(r) \left(p_1(r) g_1'(r)+q_1'(r) \phi'(r)\right)+r f_1(r) \phi''(r) \Big) \bigg] \nonumber\\
& -B(r) \phi'(r) \bigg[-4 r f_1(r) B'(r) \nonumber\\
& \qquad\qquad\qquad +r \Big(4 q_1(r) A'(r) B'(r)+r \phi'(r) A''(r) \nonumber\\
& \qquad\qquad\qquad\qquad +A'(r) \big(4 \phi'(r)+r \phi'(r)^2+2 r \phi''(r)\big)\Big) \nonumber\\
& \qquad\qquad\qquad +A(r) \Big( -4 r g_1(r) p_1(r) B'(r)+4 q_1(r) B'(r)+2 \phi'(r) \nonumber\\
& \qquad\qquad\qquad \qquad \qquad +2 r \phi'(r)^2+4 r \phi''(r)+3 r^2 \phi'(r) \phi''(r) \Big) \bigg] \nonumber\\
& -B(r)^2 \bigg[ \left(1+p_1(r)^2-q_1(r)^2\right) A'(r) B'(r)+2 r g_1(r) p_1(r) A'(r) \phi'(r) \nonumber\\
& \qquad\qquad -4 q_1(r) A'(r) \phi'(r) +2 r f_1'(r) \phi'(r)-2 r A'(r) q_1'(r) \phi'(r) \nonumber\\
& \qquad\qquad -2 r q_1(r) \phi'(r) A''(r)-2 r q_1(r) A'(r) \phi''(r) \nonumber\\
& \qquad\qquad +2 f_1(r) \left(\phi'(r)+q_1(r) B'(r) \left(1-r \phi'(r)\right)+r \phi''(r)\right) \nonumber\\
& \qquad\qquad +A(r) \Big( \left(1+p_1(r)^2+q_1(r)^2\right) B'(r) \phi'(r)-2 q_1'(r) \phi'(r) \nonumber\\
& \quad\qquad\qquad \qquad +2 p_1(r) \left(B'(r) p_1'(r)+r g_1'(r) \phi'(r)\right) -2 q_1(r) \phi''(r) \nonumber\\
& \quad\qquad\qquad \qquad +2 g_1(r) \big(r p_1'(r) \phi'(r)+p_1(r) \left(q_1(r) B'(r)+\phi'(r)+r \phi''(r)\right)\big) \Big) \bigg] \bigg\}=0\,.
\end{align}

\paragraph{Equation for $f_1(r)$.}

\begin{equation}
\frac{4 (B(r) q_1(r)+1) \left(4 \alpha_2 B(r) q_1(r) \phi'(r) + \alpha_1 r B(r)^2 - 2 \alpha_2 r \phi'(r)^2\right)}{B(r)^2}=0\,.
\end{equation}

\paragraph{Equation for $g_1(r)$.}
\begin{equation}
\frac{4 A(r) p_1(r) \left(4 \alpha_2 B(r) q_1(r) \phi'(r) + \alpha_1 r B(r)^2 - 2 \alpha_2 r \phi'(r)^2\right)}{B(r)^2}=0\,.
\end{equation}

\paragraph{Equation for $p_1(r)$.}
\begin{equation}
\begin{aligned}
\frac{4 A(r)}{B(r)^2} \bigg[ & -2 \alpha_2 \phi'(r) \left(r g_1(r) \phi'(r)-2 p_1(r) B'(r)\right) \\
& + 2 \alpha_2 B(r) \left(2 g_1(r) q_1(r) \phi'(r)+p_1(r) \left(\phi'(r)^2-2 \phi''(r)\right)\right) \\
& + \alpha_1 r B(r)^2 g_1(r) + \alpha_1 B(r)^3 p_1(r) \bigg]=0\,.
\end{aligned}
\end{equation}

\paragraph{Equation for $q_1(r)$.}
\begin{equation}
\begin{aligned}
-\frac{4}{B(r)^2} \bigg\{ & B(r)^2 \left(\alpha_1 r A'(r) - 8 \alpha_2 f_1(r) q_1(r) \phi'(r)\right) \\
& + 2 \alpha_2 B(r) \phi'(r) \left(2 q_1(r) A'(r) + f_1(r) \left(r \phi'(r)-2\right)\right) \\
& - 2 \alpha_2 r A'(r) \phi'(r)^2 - \alpha_1 r B(r)^3 f_1(r) \\
& + A(r) \bigg[ -2 \alpha_2 B(r) \phi'(r) \left(2 g_1(r) p_1(r)+q_1(r) \phi'(r)\right) \\
& \qquad\qquad + \alpha_1 B(r)^3 q_1(r) + \alpha_1 B(r)^2 - 2 \alpha_2 \phi'(r)^2 \bigg] \bigg\}=0\,.
\end{aligned}
\end{equation}

\bibliography{citationlist.bib}

@article{burgess2004quantum,
  title={Quantum gravity in everyday life: General relativity as an effective field theory},
  author={Burgess, C. P.},
  journal={Living Reviews in Relativity},
  volume={7},
  number={1},
  pages={5},
  year={2004},
  publisher={Springer}
}

@article{stelle1977renormalization,
  title={Renormalization of higher-derivative quantum gravity},
  author={Stelle, K. S.},
  journal={Physical Review D},
  volume={16},
  number={4},
  pages={953},
  year={1977},
  publisher={APS}
}

@article{starobinsky1980new,
  title={A new type of isotropic cosmological models without singularity},
  author={Starobinsky, A. A.},
  journal={Physics Letters B},
  volume={91},
  number={1},
  pages={99--102},
  year={1980},
  publisher={Elsevier}
}

@article{planck2018results,
  title={Planck 2018 results. VI. Cosmological parameters},
  author={{Planck Collaboration}},
  journal={Astronomy \& Astrophysics},
  volume={641},
  pages={A6},
  year={2020},
  publisher={EDP Sciences}
}

@article{padmanabhan2013lanczos,
  title={Lanczos--Lovelock models of gravity},
  author={Padmanabhan, T. and Kothawala, D.},
  journal={Physics Reports},
  volume={531},
  number={3},
  pages={115--171},
  year={2013},
  publisher={Elsevier}
}

@article{iyer1994some,
  title={Some properties of Noether charge and a proposal for dynamical black hole entropy},
  author={Iyer, V. and Wald, R. M.},
  journal={Physical Review D},
  volume={50},
  number={2},
  pages={846},
  year={1994},
  publisher={APS}
}

@article{brigante2008viscosity,
  title={The viscosity bound and causality violation},
  author={Brigante, M. and Liu, H. and Myers, R. C. and Shenker, S. and Yaida, S.},
  journal={Physical Review Letters},
  volume={100},
  number={19},
  pages={191601},
  year={2008},
  publisher={APS}
}

@article{dong2014holographic,
  title={Holographic entanglement entropy for general higher derivative gravity},
  author={Dong, X.},
  journal={Journal of High Energy Physics},
  volume={2014},
  number={1},
  pages={44},
  year={2014},
  publisher={Springer}
}

@article{hehl1976general,
  title={General relativity with spin and torsion: Foundations and prospects},
  author={Hehl, F. W. and von der Heyde, P. and Kerlick, G. D. and Nester, J. M.},
  journal={Reviews of Modern Physics},
  volume={48},
  number={3},
  pages={393},
  year={1976},
  publisher={APS}
}

@article{hehl1995metric,
  title={Metric-affine gauge theory of gravity: field equations, Noether identities, world spinors, and breaking of dilation invariance},
  author={Hehl, F. W. and McCrea, J. D. and Mielke, E. W. and Nester, J. M.},
  journal={Physics Reports},
  volume={258},
  number={1-2},
  pages={1--171},
  year={1995},
  publisher={Elsevier}
}

@book{blagojevic2012gauge,
  title={Gauge Theories of Gravitation},
  editor={Blagojevi{\'c}, M. and Hehl, F. W.},
  year={2012},
  publisher={Imperial College Press}
}

@article{castellani2017palatini,
  title={Palatini-Lovelock-Cartan theories},
  author={Castellani, L. and de Souza, M. C. C. and D'Auria, R.},
  journal={arXiv preprint arXiv:1704.03212},
  year={2017}
}

@article{baekler2011gravity,
  title={Gravity with torsion and nonmetricity},
  author={Baekler, P. and Hehl, F. W.},
  journal={Classical and Quantum Gravity},
  volume={28},
  number={21},
  pages={215017},
  year={2011},
  publisher={IOP Publishing}
}

@article{cvetkovic2018black,
  title={A black hole with torsion in 5D Lovelock gravity},
  author={Cvetkovi{\'c}, Branislav and Simi{\'c}, Dejan},
  journal={Classical and Quantum Gravity},
  volume={35},
  number={5},
  pages={055005},
  year={2018},
  publisher={IOP Publishing}
}

@article{canfora2014exact,
  title={Exact solutions with torsion in 5D Lovelock-Cartan theory},
  author={Canfora, F. and Giacomini, A. and Willison, S.},
  journal={Physical Review D},
  volume={88},
  number={8},
  pages={084055},
  year={2013},
  publisher={APS}
}

@article{oliva2011static,
  title={Static black holes in vacuum for Chern--Simons gravity in D= 5},
  author={Oliva, J. and Ray, S.},
  journal={Classical and Quantum Gravity},
  volume={28},
  number={17},
  pages={175009},
  year={2011},
  publisher={IOP Publishing}
}

@article{iosifidis2021riemann,
  title={Riemann tensor and Gauss--Bonnet density in metric-affine cosmology},
  author={Iosifidis, D.},
  journal={Classical and Quantum Gravity},
  volume={38},
  number={19},
  pages={195002},
  year={2021},
  publisher={IOP Publishing}
}

@article{glavan2020einstein,
  title={Einstein-Gauss-Bonnet gravity in four-dimensional spacetime},
  author={Glavan, D. and Lin, C.},
  journal={Physical Review Letters},
  volume={124},
  number={8},
  pages={081301},
  year={2020},
  publisher={APS}
}

@article{lu2020horndeski,
  title={Horndeski gravity as D$\to$ 4 limit of Gauss-Bonnet},
  author={L{\"u}, H. and Pang, Y.},
  journal={Physics Letters B},
  volume={809},
  pages={135717},
  year={2020},
  publisher={Elsevier}
}

@article{kobayashi2020effective,
  title={Effective field theory of 4D Einstein-Gauss-Bonnet gravity},
  author={Kobayashi, T.},
  journal={arXiv preprint arXiv:2003.12771},
  year={2020}
}

@article{fernandes2020derivation,
  title={Derivation of regularized field equations for the Einstein-Gauss-Bonnet theory in four dimensions},
  author={Fernandes, Pedro GS and Carrilho, Pedro and Clifton, Timothy and Mulryne, David J},
  journal={Physical Review D},
  volume={102},
  number={2},
  pages={024025},
  year={2020},
  publisher={APS}
}

@article{fernandes20224d,
  title={The 4D Einstein--Gauss--Bonnet theory of gravity: a review},
  author={Fernandes, P. G. S. and Carrilho, P. and Clifton, T. and Mulryne, D. J.},
  journal={Classical and Quantum Gravity},
  volume={39},
  number={6},
  pages={063001},
  year={2022},
  publisher={IOP Publishing}
}

@article{wei2020extended,
  title={Extended thermodynamics and microstructures of four-dimensional charged Gauss-Bonnet black hole in AdS space},
  author={Wei, S.-W. and Liu, Y.-X.},
  journal={Physical Review D},
  volume={101},
  number={10},
  pages={104018},
  year={2020},
  publisher={APS}
}

@article{haghani2020growth,
  title={Growth of matter density perturbations in 4D Einstein--Gauss--Bonnet gravity},
  author={Haghani, Z.},
  journal={Physics of the Dark Universe},
  volume={30},
  pages={100720},
  year={2020},
  publisher={Elsevier}
}

@article{gohain2024emergent,
  title={Emergent cosmology in 4D Einstein–Gauss–Bonnet theory of gravity},
  author={Gohain, M. M. and Bhuyan, K.},
  journal={Physica Scripta},
  volume={99},
  number={7},
  pages={075306},
  year={2024},
  publisher={IOP Publishing}
}

@article{khodabakhshi2024observational,
  title={Observational feasibility of 4D Einstein–Gauss–Bonnet cosmology: Bouncing and non-bouncing universes},
  author={Khodabakhshi, H. and Farhang, M. and L{\"u}, H.},
  journal={Journal of Cosmology and Astroparticle Physics},
  volume={2024},
  number={05},
  pages={024},
  year={2024},
  publisher={IOP Publishing}
}

@article{clifton2020observational,
  title={Observational constraints on the regularized 4D Einstein-Gauss-Bonnet theory of gravity},
  author={Clifton, T. and Carrilho, P. and Fernandes, P. G. S. and Mulryne, D. J.},
  journal={Physical Review D},
  volume={102},
  number={8},
  pages={084005},
  year={2020},
  publisher={APS}
}

@article{cvetkovic2019entropy,
  title={Entropy in Poincar{\'e} gauge theory: Hamiltonian approach},
  author={Cvetkovi{\'c}, B. and Blagojevi{\'c}, M.},
  journal={Physical Review D},
  volume={99},
  number={10},
  pages={104058},
  year={2019},
  publisher={APS}
}

@article{blagojevic2022entropy,
  title={Entropy of Kerr-Newman-AdS black holes with torsion},
  author={Blagojevi{\'c}, M. and Cvetkovi{\'c}, B.},
  journal={Physical Review D},
  volume={105},
  number={10},
  pages={104014},
  year={2022},
  publisher={APS}
}

@article{blagojevic2006black,
  title={Black hole entropy in 3D gravity with torsion},
  author={Blagojevi{\'c}, M. and Cvetkovi{\'c}, B.},
  journal={Classical and Quantum Gravity},
  volume={23},
  number={14},
  pages={4781},
  year={2006},
  publisher={IOP Publishing}
}

@article{chakrabarty2018different,
  title={Different types of torsion and their effect on the dynamics of fields},
  author={Chakrabarty, Subhasish and Lahiri, Amitabha},
  journal={The European Physical Journal Plus},
  volume={133},
  number={6},
  pages={242},
  year={2018},
  publisher={Springer}
}

@article{nieh1982quantized,
  title={Quantized Dirac field in curved Riemann-Cartan background. I. Symmetry properties, Green's function},
  author={Nieh, HT and Yan, ML},
  journal={Annals of Physics},
  volume={138},
  number={2},
  pages={237--259},
  year={1982},
  publisher={Elsevier}
}

@article{chen2018new,
  title={A new asymptotical flat and spherically symmetric solution in the generalized Einstein--Cartan--Kibble--Sciama gravity and gravitational lensing},
  author={Chen, Songbai and Zhang, Lu and Jing, Jiliang},
  journal={The European Physical Journal C},
  volume={78},
  number={11},
  pages={981},
  year={2018},
  publisher={Springer}
}

@article{cembranos2017new,
  title={New torsion black hole solutions in Poincar{\'e} gauge theory},
  author={Cembranos, Jose AR and Valcarcel, Jorge Gigante},
  journal={Journal of Cosmology and Astroparticle Physics},
  volume={2017},
  number={01},
  pages={014},
  year={2017},
  publisher={IOP Publishing}
}

@article{mcnutt2024symmetries,
  title={Symmetries in Riemann-Cartan Geometries},
  author={McNutt, David D and Coley, Alan A and Hoogen, Robert J},
  journal={arXiv preprint arXiv:2401.00780},
  year={2024}
}

@incollection{obukhov2023poincare,
  title={Poincar{\'e} gauge gravity primer},
  author={Obukhov, Yuri N},
  booktitle={Modified and Quantum Gravity: From Theory to Experimental Searches on All Scales},
  pages={105--143},
  year={2023},
  publisher={Springer}
}

@article{mardones1991lovelock,
  title={Lovelock-Cartan theory of gravity},
  author={Mardones, Alejandro and Zanelli, Jorge},
  journal={Classical and Quantum Gravity},
  volume={8},
  number={8},
  pages={1545},
  year={1991},
  publisher={IOP Publishing}
}

@article{nieh1980gauss,
  title={Gauss--Bonnet and Bianchi identities in Riemann--Cartan type gravitational theories},
  author={Nieh, HT},
  journal={J. Math. Phys.(NY);(United States)},
  volume={21},
  number={6},
  year={1980},
  publisher={Institute for Theoretical Physics, State University of New York at Stony~…}
}

@article{huang2009sitter,
  title={de Sitter spacetimes with torsion in the model of de Sitter gauge theory of gravity},
  author={Huang, Chao-Guang and Ma, Meng-Sen},
  journal={Physical Review D—Particles, Fields, Gravitation, and Cosmology},
  volume={80},
  number={8},
  pages={084033},
  year={2009},
  publisher={APS}
}

@article{iosifidis2019scale,
  title={Scale transformations in metric-affine geometry},
  author={Iosifidis, Damianos and Koivisto, Tomi},
  journal={Universe},
  volume={5},
  number={3},
  pages={82},
  year={2019},
  publisher={MDPI}
}
\end{document}